\documentclass[11pt]{JHEP3}

\usepackage{amsmath, axodraw, subfigure, bm, cite, graphicx, slashed,
  varioref, multirow, booktabs}

\newcommand\ie{\textit{i.e.}}  \newcommand\eg{\textit{e.g.}}
\newcommand\cf{\textit{cf.}}  
 \newcommand{\etc}{\textit{etc.}}
  
\newcommand{\mhvbar}{$\overline{\text{MHV}}$}
\DeclareMathOperator{\tr}{tr} \renewcommand{\vec}[1]{{\mathbf{#1}}}
\newcommand\mdot{\!\cdot\!}
\newcommand\gA{\mathcal{A}} \newcommand\gB{\mathcal{B}}
\newcommand\gD{\mathcal{D}} \newcommand\gF{\mathcal{F}}

\newcommand\mpp{\text{$-$$+$$+$}} \newcommand\ppm{\text{$+$$+$$-$}}
 \newcommand\mmp{\text{$-$$-$$+$}}

\newcommand\fourplus{\text{$+$$+$$+$$+$}}
\newcommand\NC{N_{\text{C}}}

\newcommand{\di}[2]{\delta_{i_{#1}}^{\bar\imath_{#2}}}

\title{The MHV QCD Lagrangian}

\author{ James~H.~Ettle, Tim~R.~Morris and
  Zhiguang~Xiao\\
  School of Physics and Astronomy,  University of Southampton\\
  Highfield, Southampton, SO17 1BJ, U.K.\\
  E-mails: \email{jhe@phys.soton.ac.uk},
  \email{T.R.Morris@soton.ac.uk}, \email{z.g.xiao@phys.soton.ac.uk} }

\preprint{SHEP 08-20}
\abstract{ We perform a canonical change of the field variables of
  light-cone gauge massless QCD to obtain a lagrangian whose terms are
  proportional up to polarisation factors to MHV amplitudes and
  continued off shell by the CSW prescription.  We solve for this
  transformation as a series expansion to all orders in the new
  fields, and use this to prove that the resulting vertices are indeed
  MHV vertices as claimed. We also demonstrate how this works
  explicitly for the vertices with: two quarks and two gluons, four
  quarks, and a particular helicity configuration of two quarks and
  three gluons.  Finally, we generalise the construction to massive
  QCD.  }

\maketitle
\begin{document}

\section{Introduction}
\label{sec:intro}

The calculation of scattering amplitudes for perturbative gauge
theories has seen several advancements recently, both in terms of
known analytic expressions and the computational techniques
underpinning them.  This is largely due to the fact that these
theories appear to have a much simpler structure than the traditional
Feynman graph computations would suggest. Parke and Taylor obtained
very compact expressions for the MHV (and \mhvbar) amplitudes
\cite{Parke:1986gb,Berends:1988zn,Parke:1989vn}.  Later, Cachazo,
Svr\v cek and Witten (CSW) discovered \cite{Cachazo:2004kj} the MHV
rules: inspired by twistor-string theory\cite{Witten:2003nn}, they
observed that tree-level amplitudes of arbitrary helicity
configurations could be assembled from Parke--Taylor amplitudes,
continued off shell by a particular prescription, and connected by
scalar propagators.  An important related development came in the form
of the BCFW recurrence relations between $S$-matrix elements,
discovered by Britto, Cachazo, Feng and Witten, described in
\cite{Britto:2004ap} and proved in \cite{Britto:2005fq}.  These can be
viewed as an indirect proof of CSW's MHV rules (proved directly in
\cite{Risager:2005vk}).  In \cite{Wu:2004jxa }, Wu and Zhu extended
the MHV rules to calculate the amplitudes with quarks and/or gluinos.

The MHV rules are tantalisingly suggestive of a field theory and have
since seen extensive application at the tree level (see \eg\
\cite{Cachazo:2004kj, Georgiou:2004wu, Georgiou:2004by, Zhu:2004kr,
  Wu:2004fb, Kosower:2004yz }) and also at one loop (such as in \cite{
  Brandhuber:2004yw, Bedford:2004py, Bedford:2004nh,
  Brandhuber:2005kd, Quigley:2004pw, Brandhuber:2006bf,
  Brandhuber:2007vm}).  This was the inspiration that kick-started the
Canonical MHV Lagrangian with Mansfield's paper
\cite{Mansfield:2005yd} in 2005. This demonstrated that a particular
canonical transformation to the field variables of the light-cone
gauge Yang--Mills resulted in another lagrangian whose vertices were
Parke--Taylor MHV amplitudes, with the CSW prescription arising
naturally from the gauge choice. (For a different approach to this,
see ref.\ \cite{Gorsky:2005sf}.)

The concrete form of this transformation was obtained in
\cite{Ettle:2006bw}, where the Parke--Taylor form of the vertices was
also shown explicitly for up to five gluons. This work was later
extended in \cite{Ettle:2007qc}, which gave a prescription for
dimensional regularisation of the Canonical MHV Lagrangian. It was
also seen that the `missing' (parts of) amplitudes --- those that
cannot be constructed using the MHV rules, such as $(\ppm)$ in $(2,2)$
signature or one-loop all-$+$ helicity --- can be recovered by
realising that the field transformation itself provided extra
`completion' vertices.  Although in ref. \cite{Ettle:2007qc} this was
only shown for the $(\fourplus)$ amplitude, it is clear from the
details of the calculation that together with the completion vertices
all that has happened by the change of variables is that the various
contributions from the light-cone Feynman diagrams have been
algebraically rearranged and thus we have no doubt that all of the
original light-cone Yang--Mills theory can be recovered in this way.

Field transformation approaches to deriving MHV amplitudes have also
been applied to other theories. For example, in ref.\
\cite{Feng:2006yy} MHV lagrangian techniques were developed for ${\cal
  N}=4$ supersymmetric Yang--Mills; and in ref.\ \cite{Ananth:2007zy},
the authors use a field transformation to make manifest the KLT
relations \cite{Kawai:1985xq} for the three- and four-graviton
vertices at the action level. In ref.\ \cite{Brandhuber:2006bf}, the
authors subject light-cone Yang--Mills to a holomorphic transformation
with a non-trivial jacobian, and argue that this gives rise to the
all-$+$ amplitudes.  Later in \cite{Brandhuber:2007vm}, the Mansfield
transformation is combined with a Lorentz-violating two-point one-loop
counterterm which the authors argue generates the infinite sequence of
one-loop all-$+$ scattering amplitudes.  An approach complementary to
the Canonical MHV Lagrangian was developed in refs.\
\cite{Boels:2007qn, Boels:2007gv}, which makes the MHV rules manifest
by continuing Yang--Mills theory to twistor space and applying a
particular gauge fixing.  This was extended to include massive scalars
for both this approach and space-time canonical transformation
techniques in \cite{Boels:2007pj, Boels:2008ef}, where in the latter
it was shown that the field transformations induced by these
techniques are identical.

One of our main goals in this paper is to develop the ideas sketched
in section 3 of \cite{Mansfield:2005yd} and build upon the work of
\cite{Ettle:2006bw}.  Specifically, this is to extend the proof of a
canonical MHV lagrangian for pure Yang--Mills into one for massless
QCD with quarks in the fundamental representation --- that is, to
obtain a lagrangian for QCD whose Feynman rules make the MHV rules for
that theory manifest.  (We will also take the opportunity
to fill a small gap in the original proof \cite{Mansfield:2005yd} for
Yang--Mills, that appears once one realises that there are extra
completion vertices that in general need to be taken into account.)
Here, we thus demonstrate that the MHV rules for massless quarks have
a field-theoretic origin lying within the context of our earlier work,
and thus also provide the appropriate completion vertices and the
starting point for dimensional regularisation and generalisation to
other fields.  We will also apply our methodology to the case where
the quarks are massive, and use it to obtain the appropriate
generalisation of the MHV rules in this scenario.

The structure of the rest of this paper is as follows.  In section
\ref{sec:transf}, we explain the notation and conventions we will be
using. We obtain the QCD action in light-cone gauge, and then proceed
to construct the field transformation that leads to the MHV lagrangian
for massless QCD.  The proof of this follows almost immediately
\cite{Mansfield:2005yd}. We then examine the form of the MHV
lagrangian in detail, giving the precise correspondence between the
its vertices and the MHV amplitudes.  We demonstrate in section
\ref{sec:examples} that this lagrangian does indeed contain vertices
corresponding to the expressions found in the literature for MHV
amplitudes containing quarks for a few low-order cases. In section
\ref{sec:masses}, we show how the quarks may be given masses within
the MHV lagrangian mechanism.  Finally in section \ref{sec:conclusion}
we draw our conclusions and indicate directions for future research.

\section{Light-cone QCD and the Transformation}
\label{sec:transf}

In this section we will explain the conventions and notation used in
this paper; perform light-cone gauge fixing of the massless QCD
lagrangian; obtain the field transformation that gives the Canonical
MHV Lagrangian for massless QCD; and finally make concrete the
relationship between its vertices and the MHV amplitudes.

\subsection{Light-cone coordinates}
\label{ssec:lightcone-coords}
We follow closely the conventions laid out in \cite{Ettle:2007qc}.
The problem is adapted to a co-ordinate system defined by
\begin{equation}
  x^0 = \tfrac1{\sqrt 2}(t-x^3), \quad
  x^{\bar 0} = \tfrac1{\sqrt 2}(t+x^3), \quad
  z = \tfrac1{\sqrt 2}(x^1+ix^2), \quad
  \bar z = \tfrac1{\sqrt 2}(x^1-ix^2).
  \label{eq:lc-coords}
\end{equation}
Note here the presence of the $1/\sqrt{2}$ factors that preserve the
normalisation of the volume form. It is useful to employ a compact
notation for the components of 1-forms in these co-ordinates, for
which we write $(p_0, p_{\bar 0}, p_z, p_{\bar z}) \equiv (\check{p},
\hat p, p, \bar p)$; for momenta labelled by a number, we write that
number with a decoration, for example the $n^{\rm th}$ external
momentum has components $(\check{n}, \hat n, \tilde n, \bar n)$. In
these co-ordinates and with this notation, the Lorentz invariant reads
\begin{equation}
  A \cdot B = \check{A}\,\hat B + \hat A\,\check{B} - A \bar B - \bar A B.
  \label{eq:lc-invariant}
\end{equation}
We will also make extensive use of the following bilinears:
\begin{equation}
  \label{eq:lc-bilinears}
  (1\:2) := \hat 1 \tilde 2 - \hat 2 \tilde 1, \quad
  \{1\:2\} := \hat 1 \bar 2 - \hat 2 \bar 1.
\end{equation}
These can be expressed in terms of the angle and square brackets found
in the literature by considering the bispinor representation of a
4-vector $p$
\begin{equation}
  p \cdot \bar\sigma = \sqrt 2 \begin{pmatrix} \check p & -p \\
    - \bar p & \hat p \end{pmatrix},
\end{equation}
where we define $\sigma^\mu = (1, \pmb\sigma)$ and $\bar\sigma^\mu =
(1, -\pmb\sigma)$, and $\pmb\sigma$ is the 3-vector of Pauli matrices.
For null $p$, this factorises as
$(p\cdot\bar\sigma)_{\alpha\dot\alpha}=\lambda_\alpha
\tilde\lambda_{\dot\alpha}$ where we can choose
\begin{equation}
  \label{eq:lc-spinors}
  \lambda_\alpha = 2^{1/4} \begin{pmatrix} -p/\sqrt{\hat p} \\ \sqrt{\hat p}
  \end{pmatrix}
  \quad\text{and}\quad
  \tilde\lambda_{\dot\alpha} = 2^{1/4} \begin{pmatrix} -\bar p/\sqrt{\hat p} \\ \sqrt{\hat p}
  \end{pmatrix}.
\end{equation}
We note here that this choice of spinors for off-shell $p$ coincides
with the CSW prescription \cite{Cachazo:2004kj} when using
reference spinors $\nu = \tilde\nu = (2^{1/4},0)^{\rm T}$
\cite{Mansfield:2005yd, Ettle:2006bw}.  Hence the spinor brackets can
be expressed as
\begin{equation}
  \label{eq:lc-spinorbrackets}
  \langle 1\:2 \rangle := \epsilon^{\alpha\beta}
  \lambda_{1\alpha} \lambda_{2\beta}
  = \sqrt{2} \frac{(1\:2)}{\sqrt{\hat 1 \hat 2}}
  \quad\text{and}\quad [ 1\:2 ] := \epsilon^{\dot\alpha\dot\beta}
  \lambda_{1\dot\alpha} \lambda_{2\dot\beta}
  = \sqrt{2} \frac{\{1\:2\}}{\sqrt{\hat 1 \hat 2}}.
\end{equation}

\subsection{The light-cone Dirac equation}
\label{sec:lcdirac}

We use the chiral Weyl representation of the Dirac matrices
\begin{equation}
  \gamma^\mu = \begin{pmatrix} 0 & \sigma^\mu \\
    \bar\sigma^\mu & 0 \end{pmatrix}.
\end{equation}
A Dirac spinor $\psi$ may be decomposed into its left- and
right-handed Weyl components $\bar\omega^{\dot\alpha}$ and
$\varphi_\alpha$ respectively as
\begin{equation}
  \psi = \begin{pmatrix} \bar\omega^{\dot\alpha} \\ \varphi_\alpha \end{pmatrix}.
\end{equation}
For massless quarks, we solve the Dirac equation $\slashed p \psi = 0$
and obtain the following polarisation spinors:
\begin{equation*}
  \bar u^+(p) \equiv (\bar\varphi(p), 0)
  \quad\text{and}\quad
  \bar u^-(p) \equiv (0, \omega(p))
\end{equation*}
for positive- and negative-helicity \emph{out-going} quarks,
respectively; and similarly
\begin{equation*}
  v^+(p) \equiv \begin{pmatrix} \bar\omega(p) \\ 0 \end{pmatrix}
  \quad\text{and}\quad
  v^-(p) \equiv \begin{pmatrix} 0 \\ \varphi(p) \end{pmatrix}
\end{equation*}
for \emph{out-going} antiquarks. Here we have used the following
definitions for the Weyl spinors:
\begin{align}
  \label{eq:poln-aqm}
  \varphi(p) &= 2^{1/4} \begin{pmatrix}
    - p/\sqrt{\hat p} \\ \sqrt{\hat p} \end{pmatrix}, \\
  \label{eq:poln-aqp}
  \bar\omega(p) &= 2^{1/4} \begin{pmatrix} \sqrt{\hat p} \\
    \bar p/\sqrt{\hat p} \end{pmatrix},
  \\
  \label{eq:poln-qp}
  \bar\varphi(p) &= 2^{1/4} ( {-\bar p/\sqrt{\hat
      p}}, \sqrt{\hat p}), \\
  \label{eq:poln-qm}
  \omega(p) &= 2^{1/4} ( \sqrt{\hat p}, {p/\sqrt{\hat p}} ),
\end{align}
chosen so that external fermion states have the conventional
phenomenologists' normalisation of $u^\dagger(p) u(p) = 2p^t$.

\subsection{Massless light-cone QCD}
\label{ssec:lcqcd}

We will study a massless QCD theory with a ${\rm SU}(N_{\rm C})$ gauge
theory. Its action is
\begin{equation}
  \label{eq:lcqcd}
  S_\text{QCD} = \int d^4x \: \bar\psi \: i \slashed \gD \: \psi
  + \frac 1{2g^2} \int d^4x \: \tr \gF^{\mu\nu} \gF_{\mu\nu}
\end{equation}
where
\begin{equation}
  \psi = (\alpha^+, \beta^+, \beta^-, \alpha^-)^{\rm T}
  \quad\text{and}\quad
  \bar\psi = (\bar\beta^+, \bar\alpha^+, \bar\alpha^-, \bar\beta^-)
\end{equation}
are the quark field and conjugate in the fundamental representation of
${\rm SU}(N_{\rm C})$. Note here that the superscripts $\pm$ on the
spinor components denote the physical helicity for \emph{out-going}
particles, and the bar denotes a field in the conjugate
representation; that $\bar\alpha^+ = (\alpha^-)^*$ should be
understood, and similarly for $\beta$.  We further define
\begin{equation}
  \label{eq:gauge-conv}
  \gF_{\mu\nu} = [\gD_\mu, \gD_\nu], \quad
  \gD_\mu = \partial_\mu + \gA_\mu, \quad
  \gA_\mu = - \frac{ig}{\sqrt 2} A^a_\mu T^a.
\end{equation}
Our gauge group generators are normalised according to
\begin{equation} [T^a, T^b] = i \sqrt 2 f^{abc} T^c, \quad \tr (T^a
  T^b) = \delta^{ab}.
\end{equation}

As in \cite{Mansfield:2005yd,Ettle:2006bw,Ettle:2007qc}, we quantise
the theory on surfaces $\Sigma$ of constant $x^0$, \ie\ those with
normal $\mu = (1,0,0,1)/\sqrt 2$ in Minkowski co-ordinates. We choose
an axial gauge $\mu \cdot \gA = \hat\gA = 0$, for which the
Faddeev--Popov ghosts are completely decoupled. The lagrangian density
is quadratic in the $\check\gA$ field; furthermore, only the fermion
field components $\bar\alpha^-$, $\bar\alpha^+$, $\alpha^-$ and
$\alpha^+$ are dynamical (inasmuch as they are the only ones to occur
in the lagrangian density in terms with $\check\partial$ quantisation
`time' derivatives). Thus we integrate $\check \gA$ and the remaining
fermion components out of the partition function and obtain the
gauge-fixed action
\begin{equation}
  \label{eq:lcqcd-action}
  \begin{split}
    S_\text{LCQCD} = \frac 4{g^2}\int dx^0 ( &L^{-+} + L^{-++} +
    L^{--+}
    + L^{--++} + \\
    &L^{\bar\psi\psi} + L^{\bar\psi+\psi} + L^{\bar\psi-\psi} +
    L^{\bar\psi+-\psi} + L^{\bar\psi\psi\bar\psi\psi}),
  \end{split}
\end{equation}
where
\begin{align}
  \label{eq:lcqcd-mp}
  L^{-+} &= \phantom{-}{\rm tr}\int_\Sigma d^3{\bf x}\: {\bar
    \gA}(\check\partial\hat\partial-
  \partial\bar\partial)\gA, \\
  \label{eq:lcqcd-mpp} {L}^{\mpp}&=-{\rm tr}\int_\Sigma d^3{\bf x}\:
  ({\bar\partial}{\hat\partial}^{-1} { \gA})\:
  [{  \gA},{\hat\partial} {\bar\gA}], \\
  \label{eq:lcqcd-mmp} {L}^{\mmp}&=-{\rm tr}\int_\Sigma d^3{\bf x}\:
  [{\bar \gA},{\hat\partial} { \gA}]\:
  ({  \partial}{\hat\partial}^{-1} {\bar \gA}), \\
  \label{eq:lcqcd-mmpp} {L}^{--++}&=-{\rm tr}\int_\Sigma d^3{\bf x}\:
  [{\bar \gA },{\hat\partial} { \gA }]\:{\hat\partial}^{-2}\: [{ \gA
  },{\hat\partial} {\bar \gA }]
\end{align}
is the pure Yang--Mills sector of the theory, and the terms involving
the fermions are \newcommand\fintl{\frac{i g^2}{\sqrt 8}\int_\Sigma
  d^3{\bf x}\: \Bigl\{}
\begin{align}
  \label{eq:lcqcd-qbarq}
  L^{\bar\psi \psi} &= \phantom{-} \fintl \bar\alpha^+ (\check\partial
  - \omega) \alpha^- + \bar\alpha^-
  (\check\partial - \omega) \alpha^+ \Bigr\}, \\
  \label{eq:lcqcd-qbarpq}
  \begin{split}
    L^{\bar\psi+\psi} &= - \fintl \bar\alpha^+ \:
    \bar\partial\hat\partial^{-1} (\gA \, \alpha^-) + \bar\alpha^- \gA
    \: \bar\partial
    \hat\partial^{-1} \alpha^+ \\
    &\hphantom{=-\fintl} - \bar\alpha^+(\bar\partial \hat\partial^{-1}
    \gA) \alpha^- - \bar\alpha^- (\bar\partial \hat\partial^{-1} \gA)
    \alpha^+ \Bigr\},
  \end{split} \\
  \label{eq:lcqcd-qbarmq}
  \begin{split}
    L^{\bar\psi-\psi} &= - \fintl \bar\alpha^+
    \bar\gA\:\partial\hat\partial^{-1}\alpha^- +
    \bar\alpha^-\partial\hat\partial^{-1}(\bar\gA \, \alpha^+) \\
    &\hphantom{=-\fintl} - \bar\alpha^+(\partial \hat\partial^{-1}
    \bar\gA) \alpha^- - \bar\alpha^- (\partial \hat\partial^{-1}
    \bar\gA) \alpha^+ \Bigr\},
  \end{split} \\
  \label{eq:lcqcd-qbarpmq}
  \begin{split}
    L^{\bar\psi+-\psi} &= - \fintl \bar\alpha^+ \bar\gA
    \hat\partial^{-1} (\gA \alpha^-)
    + \bar\alpha^- \gA \hat\partial^{-1} (\bar\gA \alpha^+) \\
    &\hphantom{=-\fintl} + \bar\alpha^+ \hat\partial^{-2}
    (\hat\partial \gA \bar\gA - \gA \hat\partial \bar\gA) \alpha^- +
    \bar\alpha^+ \hat\partial^{-2} (\hat\partial \bar\gA \gA
    - \bar\gA \hat\partial \gA) \alpha^- \\
    &\hphantom{=-\fintl} + \bar\alpha^- \hat\partial^{-2}
    (\hat\partial \gA \bar\gA - \gA \hat\partial \bar\gA) \alpha^+ +
    \bar\alpha^- \hat\partial^{-2} (\hat\partial \bar\gA \gA - \bar\gA
    \hat\partial
    \gA) \alpha^+ \Bigr\}, \end{split} \\
  \label{eq:lcqcd-qbarqqbarq}
  L^{\bar\psi\psi\bar\psi\psi} &= \frac{g^4}{16} \int_\Sigma d^3{\bf
    x}\: j^a \hat\partial^{-2} j^a, \quad j^a = \sqrt 2 (\bar\alpha^+
  T^a \alpha^- + \bar\alpha^- T^a \alpha^+).
\end{align}
Note that in \eqref{eq:lcqcd-qbarq} we define the differential
operator $\omega := \partial \bar\partial / \hat\partial$.

\subsection{Structure of the canonical transformation}
\label{ssec:transform}

Let us now construct the field transformation that results in a MHV
lagrangian for massless QCD. We label the new algebra-valued gauge
fields $\gB$ and $\bar\gB$, and the new fundamental representation
fermions $\xi^+$, $\bar\xi^-$, $\xi^-$ and $\bar\xi^+$; their Lorentz
transformation properties are the same as those of the old fields of
similar decoration.  We remove terms in the light-cone lagrangian with
a $(\mpp)$ helicity structure by absorbing them into the kinetic terms
of the new fields:
\begin{align}
  \label{eq:transf-def}
  L^{-+}[\gA, \bar\gA] + L^{-++}[\gA, \bar\gA] +
  L^{\bar\psi\psi}[\alpha^\pm,\bar\alpha^\pm] +
  L^{\bar\psi+\psi}[\gA,\alpha^\pm,\bar\alpha^\pm] &= L^{-+}[\gB,
  \bar\gB] + L^{\bar\psi\psi}[\xi^\pm,\bar\xi^\pm].
\end{align}
The remaining terms in the lagrangian, \eqref{eq:lcqcd-mmp},
\eqref{eq:lcqcd-mmpp} and
\eqref{eq:lcqcd-qbarmq}--\eqref{eq:lcqcd-qbarqqbarq} form the MHV
vertices as we will show in the next section.  We note that, at least
classically, this choice of transformation seems sensible since it
maps a field theory that is free (at the tree-level of the $S$-matrix)
on the LHS of \eqref{eq:transf-def} onto a strictly free theory in the
new variables on the RHS.

We note that the canonical $(\text{co-ordinate},\text{momentum})$
pairs of the system \eqref{eq:lcqcd-action} are
\begin{equation}
  \label{eq:oldfields}
  (\gA, -\hat\partial\bar\gA),\quad
  \left(\alpha^+, -\frac{ig^2}{\sqrt 8} \bar\alpha^- \right)
  \quad\text{and}\quad
  \left(\alpha^-, -\frac{ig^2}{\sqrt 8} \bar\alpha^+ \right),
\end{equation}
and likewise for the new fields (by replacing $\gA \rightarrow \gB$
and $\alpha \rightarrow \xi$ above).  We have defined the momenta with
respect to the lagrangian of
\eqref{eq:lcqcd-mp}--\eqref{eq:lcqcd-qbarqqbarq}.  We see that up to a
constant the path integral measure
\begin{equation} {\cal D}\gA {\cal D}\bar\gA \:
  {\cal D}\alpha^+ {\cal D}\bar\alpha^- \:
  {\cal D}\alpha^- {\cal D}\bar\alpha^+
\end{equation}
is equal to the phase space measure, and thus will be preserved if the
transformation is canonical. This, and our demands on the helicity
content of the resulting lagrangian, restrict the form of the
transformation as follows.

To begin, we choose $\gA$ to be a functional of $\gB$ alone.  This,
and the canonical transformation properties of a conjugate momentum,
imply that
\begin{equation}
  \label{eq:Abar-canon}
  \hat\partial \bar\gA^a(\vec x) = \int_\Sigma d^3\vec y \left\{
    \frac{\delta\gB^b(\vec y)}{\delta\gA^a(\vec x)}
    \hat\partial \bar\gB^b(\vec y)
    - \frac{ig^2}{\sqrt 8}\left(
      \bar\xi^-(\vec y)
      \frac{\delta \xi^+(\vec y)}{\delta\gA^a(\vec x)}
      +  \bar\xi^+(\vec y) \frac{\delta \xi^-(\vec y)}{\delta\gA^a(\vec x)}
    \right)
  \right\}
\end{equation}
where all fields have the same implicit $x^0$ dependence. Note that we
take all derivatives with respect to Grassman variables as acting from
the left.  By charge conservation, and the requirement that this will
be a canonical transformation that results in a lagrangian whose
vertices have MHV helicity content, the fermion co-ordinate
transformation takes the form (which we explain below)
\begin{equation}
  \label{eq:fqnew-in-old}
  \xi^\pm(\vec x) = \int_\Sigma d^3\vec y \: R^\mp[\gA](\vec x, \vec y)
  \: \alpha^\pm(\vec y).
\end{equation}
The superscript of $R^\pm$ refers to the chirality of the Weyl spinor
from which the fermion components originate: $+$ for right-handed, $-$
for left-handed.  $R^\pm$ is a matrix-valued functional of
$\gA$. Putting additional factors of $\bar\gA$ into the RHS of
\eqref{eq:fqnew-in-old} would result in terms in the resulting
lagrangian with more than two fields of negative helicity; likewise
with extra quark fields, since charge conservation requires these to
be added in $(+-)$ helicity pairs.  The behaviour of the canonical
momentum under a canonical transformation is then fixed by
\eqref{eq:fqnew-in-old} to be
\begin{equation}
  \label{eq:fpold-in-new}
  \bar\alpha^\pm(\vec x) = \int_\Sigma d^3\vec y \:
  \bar\xi^\pm(\vec y) \: R^\pm[\gA](\vec y, \vec x),
\end{equation}
but it will also be useful to define the inverse transformations
\begin{equation}
  \label{eq:fqold-in-new}
  \alpha^\pm(\vec x) = \int_\Sigma d^3\vec y \: S^\mp[\gA](\vec x, \vec y)
  \: \xi^\pm(\vec y)
\end{equation}
as well.

At this point we can immediately read off the propagators for the new
fields from \eqref{eq:transf-def} as
\begin{equation}
  \label{eq:propagators}
  \langle \gB \bar\gB \rangle = - \frac{ig^2}{2p^2}
  \quad\text{and}\quad
  \langle \xi^- \bar\xi^+ \rangle = \langle \xi^+ \bar\xi^- \rangle = i \sqrt 2 \frac{\hat p}{p^2}.
\end{equation}
By using \eqref{eq:gauge-conv}, one obtains the canonically normalised
propagator $\langle B \bar B \rangle = i/p^2$, and indeed in practical
calculations with the MHV lagrangian it is often more convenient to
absorb powers of this factor into the lagrangian's vertices and
transformation series coefficients, as was done in
\cite{Ettle:2007qc}. For the purposes of this paper, however, we will
account for these factors at the end of the calculations we present in
the forthcoming.

\TABLE{%
  \begin{tabular}{ll}
    \toprule
    \textbf{LCQCD term} & \textbf{New field content} \\
    \midrule
    $L^\mmp$ & $\bar\gB\bar\gB\gB\cdots$,
    \quad $\bar\xi\xi\bar\gB\gB\cdots$,
    \quad $\bar\xi\xi\bar\xi\xi\gB\cdots$ \\
    $L^{--++}$ & $\bar\gB\bar\gB\gB\gB\cdots$,
    \quad $\bar\xi\xi\bar\gB\gB\cdots$,\quad
    $\bar\xi\xi\bar\xi\xi\gB\gB\cdots$ \\
    \midrule
    $L^{\bar\psi-\psi}$ & $\bar\xi\xi\bar\gB$,\quad
    $\bar\xi\xi\bar\gB\gB\cdots$,\quad $\bar\xi\xi\bar\xi\xi\gB\cdots$ \\
    $L^{\bar\psi+-\psi}$ & $\bar\xi\xi\bar\gB\gB\cdots$,\quad
    $\bar\xi\xi\bar\xi\xi\gB\cdots$ \\
    $L^{\bar\psi\psi\bar\psi\psi}$ & $\bar\xi\xi\bar\xi\xi$,\quad
    $\bar\xi\xi\bar\xi\xi\gB\cdots$ \\ 
    \bottomrule
  \end{tabular}
  \label{tbl:mhvqcd-content}
  \caption{The contents of the new vertices provided by our choice of
    field transformation. The new fermion fields, $\xi$, always occur
    in bilinear pairs and as such $\bar\xi\xi$ is the sum of a term
    containing exactly one $-$ helicity quark, and another term with
    one $-$ helicity antiquark. An ellipsis $\cdots$ denotes an
    infinite series wherein the field to its immediate left is
    repeated.}
}

If we now assume solutions for $R$ and $S$ as infinite series in
$\gA$, it is not hard to see that, upon substitution into the
non-transformation terms of the light-cone QCD lagrangian
\eqref{eq:lcqcd-mmp}, \eqref{eq:lcqcd-mmpp} and
\eqref{eq:lcqcd-qbarmq}--\eqref{eq:lcqcd-qbarqqbarq}, this choice of
transformation furnishes a set of terms with no more than two fields
of negative helicity ($\bar\gB$, $\xi^-$ and $\bar\xi^-$) in each, but
an increasing number of $\gB$, as shown in
table~\ref{tbl:mhvqcd-content}. (The number of positive-helicity quark
fields present is, of course, strictly constrained by charge
conservation.) Furthermore, inspection of these terms tells us that
the interaction part of the MHV lagrangian should be a sum of the
following two terms. Switching to momentum space on the quantisation
surface $\Sigma$, the first is the purely gluonic part
\begin{equation}
  L_\text{YM}^\text{MHV} = \frac 12 \sum_{n=3}^\infty
  \sum_{s=2}^{n} \int_{1\cdots n} V^s_{\text{YM}}(1 \cdots n)
  \tr ( \bar\gB_{\bar 1} \gB_{\bar 2} \cdots \bar \gB_{\bar s} \cdots
  \gB_{\bar n} ) \: (2\pi)^3 \delta^3({\textstyle \sum_{i=1}^n \vec p_i}).
  \label{eq:MHVL-massless-YM}
\end{equation}
Note here that numbered subscripts denote momentum arguments, the bar
denoting negation (\ie\ $\gB_{\bar\imath} := \gB(-\vec p_i)$), and our
integral short-hand is defined by
\begin{equation}
  \int_{1\cdots n} = \prod_{k=1}^n \frac 1{(2\pi)^3}
  \int d\hat k \: dk \: d\bar k.
\end{equation}
The factor of $\frac12$ above absorbs the factor of $2$ that arises
from the two possible contractions of gluon external states into the
trace.  The second term is a new fermionic part
\begin{multline}
  L_\text{F}^\text{MHV} = \sum_{n=3}^\infty \sum_{s=2}^{n-1}
  \int_{1\cdots n} \bigg\{ V^{s,-+}_{\text{F}}(1\cdots n) \: \bar
  \xi^-_{\bar 1} \gB_{\bar 2}\cdots \bar \gB_{\bar s} \cdots
  \gB_{\overline{n-1}}\xi^+_{\bar n}
  \\
  + V^{s,+-}_{\text{F}}(1\cdots n) \: \bar \xi^+_{\bar 1} \gB_{\bar 2}
  \cdots \bar\gB_{\bar s} \cdots \gB_{\overline{n-1}} \xi^-_{\bar
    n}\bigg\}
  \\
  \quad + \sum_{n=4}^\infty \sum_{s=2}^{n-2} \int_{1\cdots n} \bigg\{
  \frac 12 V^{s,+-+-}_{\text{F}}(1\cdots n) \: \bar\xi^+_{\bar 1}
  \gB_{\bar2} \cdots \gB_{\overline {s-1}} \xi^-_{\bar s}
  \bar\xi^+_{\overline {s+1}} \gB_{\overline{s+2}} \cdots
  \gB_{\overline {n-1}} \xi^-_{\bar n}
  \\
  + \frac 12 V^{s,-+-+}_{\text{F}}(1\cdots n) \: \bar\xi^-_{\bar 1}
  \gB_{\bar 2} \cdots \gB_{\overline {s-1}} \xi^+_{\bar s}\
  \bar\xi^-_{\overline{s+1}} \gB_{\overline{s+2}} \cdots
  \gB_{\overline {n-1}} \xi^+_{\bar n}
  \\
  + V^{s,++--}_{\text{F}}(1\cdots n) \: \bar\xi^+_{\bar 1} \gB_{\bar
    2} \cdots \gB_{\overline {s-1}} \xi^+_{\bar s}
  \bar\xi^-_{\overline{s+1}} \gB_{\overline{s+2}} \cdots
  \gB_{\overline {n-1}} \xi^-_{\bar n}
  \\
  + V^{s,+--+}_{\text{F}}(1\cdots n) \: \bar\xi^+_{\bar 1} \gB_{\bar
    2} \cdots \gB_{\overline {s-1}} \xi^-_{\bar s}
  \bar\xi^-_{\overline{s+1}} \gB_{\overline{s+2}} \cdots
  \gB_{\overline {n-1}} \xi^+_{\bar n} \bigg\}.
  \label{eq:MHVL-massless-q}
\end{multline}
Again, we point out the symmetry factors of $\frac12$, and that we
have absorbed the factors of $(2\pi)^3\delta^3(\sum_{i=1}^n\vec p_i)$
into $\int_{1\cdots n}$ for compactness.

Note that \eqref{eq:MHVL-massless-YM} and \eqref{eq:MHVL-massless-q}
have precisely the helicity and colour structure required to be
identified as the interaction part of a MHV lagrangian, and thus the
Feynman rules of its tree-level perturbation theory will follow the
MHV rules. We will review these MHV rules and detail their precise
correspondence with the MHV lagrangian in section
\ref{ssec:vertex-structure}, but we note for now that each term's
unique helicity and colour structure means that it is the sole
contributor to the corresponding MHV amplitude
\cite{Cachazo:2004kj,Wu:2004jxa}, thus on shell the vertex must be
that amplitude, up to polarisation factors.  Actually, this assertion
follows once we demonstrate that the extra completion vertices that
follow from the transformation \cite{Ettle:2007qc} make no
contribution, and we address this in section
\ref{ssec:offshell-proof}.  Off shell, each vertex is still the MHV
amplitude up to polarisation factors, continued off shell by the CSW
prescription. This follows essentially directly by the argument given
in ref.\ \cite{Mansfield:2005yd}, and again we sketch the proof in
section \ref{ssec:offshell-proof}.

\subsection{Explicit solutions}
\label{ssec:series-solns}

In the meantime, let us proceed by obtaining explicit solutions to the
canonical transformation.  Let us begin with \eqref{eq:transf-def}. We
write it out explicitly, making use of \eqref{eq:Abar-canon},
\eqref{eq:fqnew-in-old} and \eqref{eq:fpold-in-new} to substitute for
$\hat\partial\bar\gA$, $\bar\alpha^\pm$ and $\xi^\pm$ respectively.
For clarity, we show below only the left-handed chiral fermion
components:
\begin{multline}
  \label{eq:transf-1}\int_{\vec x \vec y} \bigl\{ \omega \gA + [\gA,
  \zeta \gA] \bigr\}^a(\vec x) \frac{\delta \gB^b(\vec y)}{\delta
    \gA^a(\vec x)} \hat\partial \bar\gB^b(\vec y) - \frac{ig^2}{\sqrt
    8} \int_{\vec x \vec y \vec z} \bigl\{ \omega \gA + [\gA, \zeta
  \gA ] \bigr\}^a(\vec z) \times \\ \Biggl\{ \bar\xi^-(\vec x)
  \frac{\delta R^-(\vec x, \vec y)}{\delta \gA^a(\vec z)}
  \alpha^+(\vec y) + \text{r.h.} \Biggr\} + \frac{ig^2}{\sqrt 8}
  \int_{\vec x \vec y} \Bigr[ \bar\xi^-(\vec x) \bigl\{ R^-(\vec
  x,\vec y)[\zeta \gA(\vec y)] \\ + \omega_y R^-(\vec x, \vec y) -
  \zeta_y [R^-(\vec x, \vec y) \gA(\vec y)] \bigr\} \alpha^+(\vec y) +
  \text{r.h.} \Bigr] \\ = \int_{\vec x} \omega \gB^a
  \hat\partial\bar\gB^a - \frac{ig^2}{\sqrt 8} \int_{\vec x \vec y}
  \bigl\{ \bar\xi^-(\vec x) \omega_x R^-(\vec x, \vec y) \alpha^+(\vec
  y) + \text{r.h.}  \bigr\}.
\end{multline}
We have adopted the convenient short-hand $\int_{\vec x \vec y \cdots}
:= \int_\Sigma d^3\vec x \: d^3 \vec y \cdots$, and defined the
differential operator $\zeta = \bar\partial / \hat\partial$.

Now recall from \cite{Ettle:2006bw} we obtained $\gA$ as a series in
$\gB$:
\begin{equation}
  \label{eq:A-series}
  \gA_1 = \sum_{n=2}^\infty \int_{2\cdots n} \Upsilon(1\cdots n)
  \gB_{\bar 2} \cdots \gB_{\bar n} \: (2\pi)^3 \delta^3({
    \textstyle \sum_{i=1}^n \vec p_i
  })
\end{equation}
with
\begin{equation}
  \label{eq:Upsilon-coeff}
  \Upsilon(1\cdots n) = (-i)^n \frac{\hat 1 \hat 3 \cdots \widehat{n-1}}
  {(2\:3) \cdots (n-1,n)}.
\end{equation}
This was obtained by solving equation (2.17) of \cite{Ettle:2006bw}:
\begin{equation}
  \label{eq:transf-orig}
  \int_{\vec x} \bigl\{ \omega \gA + [\gA, \zeta \gA]
  \bigr\}^a(\vec x) \frac{\delta \gB^b(\vec y)}{\delta \gA^a(\vec
    x)}  = \omega \gB^b(\vec y).
\end{equation}
Thus by substituting for $\gA$ with \eqref{eq:A-series}, we can
eliminate the first terms from either side of \eqref{eq:transf-1}
leaving only terms bilinear in the quark fields.  Furthermore, we can
use \eqref{eq:transf-orig} on the LHS of \eqref{eq:transf-1} to trade
the $\delta R^-/\delta \gA$ for $\delta R^-/\delta \gB$, and thus for
the left-handed parts we arrive at
\begin{multline}
  \label{eq:transf-gauge-3-lh}
  \int_{\vec x \vec y \vec z} \bar\xi^-(\vec x) \left\{ (\omega_x +
    \omega_y) R^-(\vec x, \vec y) - [\omega \gB^a(\vec z)]
    \frac{\delta R^-(\vec x, \vec y)}{\delta\gB^a(\vec z)} \right\}
  \alpha^+(\vec y) \\
  = \int_{\vec x \vec y \vec z} \bar\xi^-(\vec x) \left\{ \zeta_y
    [R^-(\vec x, \vec y) \gA(\vec y)] - R^-(\vec x, \vec y) [\zeta
    \gA(\vec y)] \right\} \alpha^+(\vec y).
\end{multline}
The same procedure yields a similar equation for the right-handed
sector:
\begin{multline}
  \label{eq:transf-gauge-3-rh}
  \int_{\vec x \vec y \vec z} \bar\xi^+(\vec x) \left\{ ( \omega_x +
    \omega_y ) R^+(\vec x, \vec y) - [ \omega \gB^a(\vec z) ]
    \frac{\delta R^+(\vec x, \vec y)}{\delta\gB^a(\vec z)} \right\}
  \alpha^-(\vec y) \\
  = \int_{\vec x \vec y \vec z} \bar\xi^+ (\vec x) \left\{ [\zeta_y
    R^+(\vec x, \vec y)] \gA(\vec y) - R^+(\vec x, \vec y) [ \zeta
    \gA(\vec y) ] \right\} \alpha^-(\vec y).
\end{multline}

Since the quark fields are arbitrary, equations
\eqref{eq:transf-gauge-3-lh} and \eqref{eq:transf-gauge-3-rh}
determine the solution for $R^\pm$ in terms of $\gB$. We switch to
quantisation surface momentum space, and postulate a series solution
of the form
\begin{equation}
  \label{eq:R-series}
  R^\pm(12) = (2\pi)^3 \delta^3(\vec p_1 + \vec p_2) + \sum_{n=3}^\infty
  \int_{3\cdots n} R^\pm (12;3\cdots n) \gB_{\bar 3} \cdots
  \gB_{\bar n} \: (2\pi)^3 \delta^3({\textstyle \sum_{i=1}^n \vec p_i})
\end{equation}
Here, momenta $1$ and $2$ are associated with the Fourier transforms
of $\vec x$ and $\vec y$, respectively, in \eqref{eq:fqnew-in-old}.
For future purposes, it will often be convenient to absorb the first
term above into the sum by defining $R^\pm(12;) = 1$.  Writing
equations \eqref{eq:transf-gauge-3-lh} and
\eqref{eq:transf-gauge-3-rh} in momentum space and using
\eqref{eq:A-series} to substitute for $\gA$ leads to the following two
recurrence relations:
\begin{multline}
  \label{eq:Rl-rr}
  R^-(12;3\cdots n) = \\ \frac{-i}{\omega_1 + \cdots + \omega_n}
  \sum_{j=2}^{n-1} \frac{\{2, P_{j+1,n}\}}{\hat2 \, \hat P_{j+1,n}} \:
  R^-(1,2\!+\!P_{j+1,n};3\dots j) \Upsilon(-,j\!+\!1, \cdots, n)
\end{multline}
and
\begin{multline}
  \label{eq:Rr-rr}
  R^+(12;3\cdots n) = \\\frac{-i}{\omega_1 + \cdots + \omega_n}
  \sum_{j=2}^{n-1} \frac{\{2, P_{j+1,n}\}}{(\hat2 \!+\! \hat
    P_{j+1,n}) \, \hat P_{j+1,n}} \: R^+(1,2\!+\!P_{j+1,n};3\dots j)
  \Upsilon(-,j\!+\!1, \cdots, n),
\end{multline}
where we define the momentum space analogue of the $\omega$ operator
as $\omega_p := p\bar p/\hat p$, $P_{ij}:=\sum_{k=i}^j p_k$, and $-$
as a momentum argument denotes the negative of the sum of the other
arguments. We notice immediately from the above that if we put
\begin{equation}
  \label{eq:Rl-coeff}
  R^-(12;3\cdots n) = -\frac{\hat1}{\hat2} R^+(12;3\cdots n)
\end{equation}
into \eqref{eq:Rl-rr}, we recover \eqref{eq:Rr-rr}; note that this
only fixes the numerator $\hat 1$ above, whereas the sign and the
denominator follow by noting that the lowest-order coefficients
$R^\pm(12) = 1$ are defined for conserved momentum (\ie\ at $\vec p_1=
-\vec p_2$). Thus, we need only solve for $R^+$.

Now one could obtain (and prove) a form for the $R^+$ coefficients by
direct iteration of \eqref{eq:Rr-rr} and induction on $n$, but in fact
it turns out that the recurrence relation \eqref{eq:Rr-rr} is nothing
other than a re-labelling of the one used in ref. \cite{Ettle:2006bw}
used to obtain the $\Upsilon$ coefficients,
\begin{equation}
  \label{eq:Upsilon-rr}
  \Upsilon(1\cdots n) = \frac{-i}{\omega_1 + \cdots + \omega_n}
  \sum_{j=2}^{n-1} \left( \frac{\bar P_{j+1,n}}{\hat P_{j+1,n}}
    - \frac{\bar P_{2j}}{\hat P_{2j}} \right)
  \Upsilon(-,2,\cdots,j) \Upsilon(-,j\!+\!1,\cdots,n),
\end{equation}
the solution to which was proved to be \eqref{eq:Upsilon-coeff}.  If
we now put
\begin{equation}
  \label{eq:Rr-coeff}
  R^+(12;3\cdots n) = \Upsilon(213\cdots n)
  = (-i)^n \frac{\hat2 \hat3 \cdots \widehat{n-1}}{(1\:3) (3\:4) \cdots (n-1,n)} 
\end{equation}
into \eqref{eq:Rr-rr} (and swap momenta $1$ and $2$), we arrive at
\eqref{eq:Upsilon-rr}, and \eqref{eq:Rr-coeff} is thereby proved.

The inverse fermion transformation, $S^\pm$, may be obtained from
$R^\pm$ in an order-by-order manner. Let us begin by writing it as
\begin{equation}
  \label{eq:S-series}
  S^\pm(12) = \sum_{n=2}^\infty \int_{3\cdots n}
  S^\pm (12;3\cdots n) \gB_{\bar 3} \cdots
  \gB_{\bar n} \: (2\pi)^3 \delta^3({\textstyle \sum_{i=1}^n \vec p_i}),
\end{equation}
where momenta $1$ and $2$ correspond to the Fourier transforms of
$\vec x$ and $\vec y$ in \eqref{eq:fqold-in-new}, and we have absorbed
the ${\cal O}({\cal B}^0)$ (\ie\ $n=2$) term into the sum by defining
$S^\pm(12;) = 1$. The $S^\pm$ coefficients satisfy the recurrence
relations
\begin{equation}
  \label{eq:Srl-rr}
  S^\pm(12;3\cdots n) =
  - \sum_{j=2}^{n-1} S^\pm(1,-;3 \dots j) R^\pm(-,2;j\!+\!1, \cdots, n).
\end{equation}
Now it is clear from \eqref{eq:Rl-coeff} that
\begin{equation}
  \label{eq:Sr-coeff}
  S^+(12;3\cdots n) = -\frac{\hat2}{\hat1} S^-(12;3\cdots n)
\end{equation}
where again the overall normalisation is fixed by the lowest order
coefficient. We state that \eqref{eq:Srl-rr} is solved by
\begin{equation}
  \label{eq:Sl-coeff}
  S^-(12;3 \cdots n) = (-i)^n
  \frac{\hat1 \hat 4 \cdots \hat n}{(3\:4) \cdots
    (n-1,n)(n\:2)}
  = \Upsilon(13\cdots n2)
\end{equation}
where in the case of $S^-(12;3)$ only the first factor in the
numerator and the last factor in the denominator are retained. The
proof is by induction on $n$. It is easy to check the initial step by
direct iteration of \eqref{eq:Srl-rr} for $n=3,4$, and the inductive
step is given in appendix \ref{ssec:S-proof}\footnote{We could, of
  course, have obtained $S^\pm$ in a manner similar to $R^\pm$, by
  deriving the recurrence relations and using \eqref{eq:Sl-coeff} to
  map to the $\Upsilon$ recurrence relation.}.

Finally, let us solve for $\bar\gA$. That the transformation is
canonical requires that
\begin{equation}
  \label{eq:canonicalRel-x}
  \begin{split}
    \int_{\vec x} \biggl\{ \tr \check\partial \gA \hat\partial \bar\gA
    &- \frac{ig^2}{\sqrt8}\int d^3 \vec x (\bar\alpha^- \check\partial
    \alpha^+ + \bar\alpha^+ \check\partial \alpha^-) \biggr\}
    \\
    &= \int_{\vec x} \biggl\{ \tr \check\partial \gB \hat\partial
    \bar\gB - \frac{ig^2}{\sqrt8}\int d^3 \vec x (\bar\xi^-
    \check\partial \xi^+ + \bar\xi^+ \check\partial \xi^-) \biggr\}
  \end{split}
\end{equation}

Now consider the functional form of $\hat\partial \bar\gA$ as given by
\eqref{eq:Abar-canon}. We can split it into two pieces,
\begin{equation}
  \label{eq:Abar-split}
  \hat\partial \bar\gA = \hat\partial \bar\gA^0 + \hat\partial
  \bar\gA^{\rm F},
\end{equation}
where the first term depends only on $\gB$ and $\bar\gB$, and the
second contains the fermion dependence. If we substitute this into
\eqref{eq:canonicalRel-x}, we see that we must have
\begin{equation}
  \tr \int_{\vec x} \check\partial \gA \hat\partial \bar\gA^0 = \tr \int_{\vec x} \check\partial \gB \hat\partial \bar\gB,
\end{equation}
similar to the case of ref.\ \cite{Ettle:2006bw}, but in $\bar\gA^0$
instead of $\bar\gA$.  Thus the first part of \eqref{eq:Abar-canon} is
taken care of if we use the pure-gauge solution for $\bar\gA^0$ found
in \cite{Ettle:2006bw}. In momentum space this is
\begin{equation}
  \label{eq:Abar0-series}
  \bar\gA^0_{1} = 
  \sum_{m=2}^\infty \sum_{s=2}^m \int_{2\cdots m}
  \frac{\hat s^2}{\hat 1^2}
  \Upsilon(1 \cdots m) \gB_{\bar2} \cdots \bar\gB_{\bar s}
  \cdots \gB_{\bar m} \: (2\pi)^3 \delta^3({\textstyle \sum_{i=1}^m \vec p_i}).
\end{equation}
From this, we immediately see that the pure-gauge MHV lagrangian of
\cite{Ettle:2006bw,Mansfield:2005yd} is recovered via the terms that
$\bar\gA^0$ contributes to $\bar\gA$ when used in $L^\mmp$ and
$L^{--++}$.

In quantisation surface momentum space, what is left of
\eqref{eq:canonicalRel-x} is
\begin{equation}
  \label{eq:barAR}
  -\tr \check\partial \gA_1 \hat 1 \bar\gA^{\rm F}_{\bar 1}
  +\frac{g^2}{\sqrt8} (\bar{\alpha}^-_1 \check \partial \alpha^+_{\bar 1}
  +\bar{\alpha}^+_1 \check \partial \alpha^-_{\bar 1}) =
  \frac{g^2}{\sqrt8} (\bar{\xi}^-_1 \check \partial \xi^+_{\bar 1}+
  \bar{\xi}^+_1 \check \partial \xi^-_{\bar 1} ).
\end{equation}
The form of \eqref{eq:Abar-canon} tells us that $\bar\gA^{\rm F}$ can
be expanded as
\begin{multline}
  \label{eq:AbarF-series}
  \bar\gA^{\text{F}}_{1} = -\frac{g^2}{\hat1 \sqrt8} \sum_\pm
  \sum_{n=3}^\infty \sum_{j=1}^{n-2} \int_{2\cdots n} K^{\pm(j)}(1
  \cdots n) \Biggl\{ \gB_{\bar 2} \cdots \gB_{\bar\jmath}
  \xi^{\mp}_{\overline{j+1}} \bar\xi^\pm_{\overline{j+2}}
  \gB_{\overline{j+3}} \cdots \gB_{\bar n} \\
  + \frac1{N_{\rm C}} \bar\xi^\pm_{\overline{j+2}}
  \gB_{\overline{j+3}} \cdots \gB_{\bar n} \gB_{\bar 2} \cdots
  \gB_{\bar\jmath} \xi^{\mp}_{\overline{j+1}} \Biggr\} (2\pi)^3
  \delta^3({\textstyle \sum_{i=1}^n \vec p_i})
\end{multline}
We substitute the expansions for $\gA$, $\bar\gA^{\rm F}$ and
$\alpha^\pm$ into \eqref{eq:barAR} and, after some careful
relabelling, match the coefficients of the strings of fields on either
side. The result is the following recurrence relation for the $K^\pm$
\footnote{Note that in the interest of clarity and to emphasise the
  origin of the relevant terms, we have \emph{not} substituted for
  $R^\pm$ and $S^\pm$ in terms of $\Upsilon$ in the forthcoming.}:
\begin{equation}
  \begin{split}
    K^{\pm(j)}(1\cdots n) =& -\sum^{j}_{r=2} \Upsilon(-,1,2,\dots, r)
    K^{\pm(j-r+1)}(- ,r+1,\dots,n)
    \\
    &-\sum^{n-1}_{l=j+2}\sum^{j}_{r=1} \Upsilon(-
    ,l+1,\dots,n,1,2,\dots,r) K^{\pm(j-r+1)}(- ,r+1,\dots,l)
    \\
    &-\sum^{n}_{l=j+2} R^\pm(j+2,-;j+3,\dots,l) S^\pm(-,
    j+1;l+1,\dots,n,1,2,\dots,j),
  \end{split}
  \label{eq:tildeKR}
\end{equation}
Note that in the case where a sum's upper limit is less than the lower
limit, the sum is taken to vanish.  Let us solve for the first few
coefficients.  For $n=3$, $j=1$, only the third term in
\eqref{eq:tildeKR} contributes:
\begin{equation}
  K^{-(1)}(123) = -S^-(32;1)= -i {\hat 3 \over (2\:3)},
\end{equation}
and for $n=4$, $j=1$, only the second and third term contribute:
\begin{equation}
  \begin{split}
    K^{-(1)}(1234) &= -\Upsilon(-,4,1)K^{-(1)}(-,2,3) -S^-(32;41)
    -R^-(3,-;4) S^-(-,2;1)
    \\
    &= -{{\hat3^2}\over(2\: 3)(3\:4)}.
  \end{split}
\end{equation}
Similarly, we can obtain next few coefficients:
\begin{align}
  K^{-(2)}(1234) &= - {{\hat3\hat{4}}\over(2\: 3)(3\:4)},
  \\
  K^{-(1)}(12345) &= i {{\hat3^2\hat4}\over(2\: 3)(3\:4)(4\:5)},
  \\
  K^{-(2)}(12345) &= i {{\hat3\hat4^2}\over(2\: 3)(3\:4)(4\:5)},
  \\
  K^{-(3)}(12345) &= i {{\hat3\hat4\hat5}\over(2\: 3)(3\:4)(4\:5)},
\end{align}
and so-on, from which we propose
\begin{equation}
  \label{eq:K--coeff}
  K^{-(j)}(1\cdots n) = -(-i)^n {\widehat{j+2}\:\hat3\hat 4\cdots
    \widehat{n-1}\over(2\:3)(3\:4)\cdots(n\!-\!1, n)}
  = - \frac{\widehat{j+2}}{\hat1} \Upsilon(1\cdots n).
\end{equation}
This can be proved by induction on $n$. The foregoing obviously
furnishes the initial step, and the inductive part follows upon
substituting \eqref{eq:K--coeff} into \eqref{eq:tildeKR} and
evaluating the sums. This is a straightforward but tedious
calculation, which we sketch in appendix \ref{ssec:K-proof}.

We could derive $K^+$ by a similar process, but there is a short-cut.
Notice that it satisfies the same recurrence relation as $K^-$, except
for the last term in \eqref{eq:tildeKR}. For this last term, we
observe that
\begin{multline}
  R^+(j+2,-;j+3,\dots,l) S^+(-,j+1;l+1,\dots,n,1,\dots,j)
  \\
  =-\frac {\widehat {j+1}}{\widehat{j+2}} R^-(j+2,-;j+3,\dots,l)
  S^-(-,j+1;l+1,\dots,n,1,\dots,j),
\end{multline}
so it is easy to see that
\begin{equation}
  \label{eq:K+-coeff}
  K^{+(j)}(12\cdots n)=
  -\frac{\widehat{j+1}}{\widehat{j+2}} K^{-(j)}(12\cdots n).
\end{equation}

\begin{figure}[h]
  \begin{align*}
    \begin{matrix}\begin{picture}(90,67)
        \SetOffset(45,42)
        \ArrowLine(0,13)(0,1)
        \DashArrowLine(0,0)(28.9778,-7.7646){2}
        \Line(12.6785,-27.1892)(0,0)
        \Line(0,0)(-28.9778,-7.7646) \BCirc(0,0){2}
        \DashCArc(0,0)(15,-160,-70){1}
        \Text(0,16)[bc]{$1^\pm$} \Text(31,-5)[tl]{$2^\pm$}
        \Text(12,-29)[tl]{$3^+$}
        \Text(-31,-5)[tr]{$n^+$}
      \end{picture}\end{matrix} &= R^\pm(21;3\cdots n)
    \\
    \begin{matrix}\begin{picture}(90,67)
        \SetOffset(45,42)
        \ArrowLine(0,1)(0,13)
        \Line(0,0)(28.9778,-7.7646)
        \Line(0,0)(-12.6785,-27.1892)
        \DashArrowLine(-28.9778,-7.7646)(0,0){2} \BCirc(0,0){2}
        \DashCArc(0,0)(15,-110,-20){1}
        \Text(0,16)[bc]{$1^\pm$} \Text(31,-5)[tl]{$2^+$}
        \Text(0,-29)[tr]{$(n\!-\!1)^+$}
        \Text(-31,-5)[tr]{$n^\pm$}
      \end{picture}\end{matrix} &= S^\pm(1n;2,\dots,n-1)
    \\
    \begin{matrix}\begin{picture}(90,67)
        \SetOffset(45,42)
        \Gluon(0,1)(0,13){2}{2}
        \Line(0,0)(28.9778,-7.7646)
        \DashArrowLine(12.6785,-27.1892)(0,0){2}
        \DashArrowLine(0,0)(-12.6785,-27.1892){2}
        \Line(0,0)(-28.9778,-7.7646) \BCirc(0,0){2}
        \DashCArc(0,0)(15,-55,-20){1}
        \DashCArc(0,0)(15,-160,-125){1}
        \Text(0,16)[bc]{$1^-$} \Text(31,-5)[tl]{$2^+$}
        \Text(2,-29)[tl]{$(j\!+\!1)^\mp$} \Text(-2,-29)[tr]{$(j\!+\!2)^\pm$}
        \Text(-31,-5)[tr]{$n^+$}
      \end{picture}\end{matrix} &= \frac{g^2}{\hat1^2 \sqrt 8}
    K^{\pm(j)}(1\cdots n)
  \end{align*}
  \label{fig:mhvqcd-completionverts-c}
  \caption{The non-trivial MHV completion vertices for massless QCD
    that involve fermions. Curly and solid, arrowless lines are $\gA$
    and $\gB$, respectively; solid lines with arrows represent
    correlation function insertions of $\alpha^\pm$ ($\bar\alpha^\pm$)
    when the arrow points outwards (inwards), and $\xi^\pm$
    ($\bar\xi^\pm$) attach to the dotted lines with the arrows
    pointing inwards (outwards).  The direction of the arrow shows
    charge flow in both cases, and the missing lines are for
    $+$-helicity gluons.  (All momenta are out-going.)  }
  \label{fig:mhvqcd-completionverts}
\end{figure}
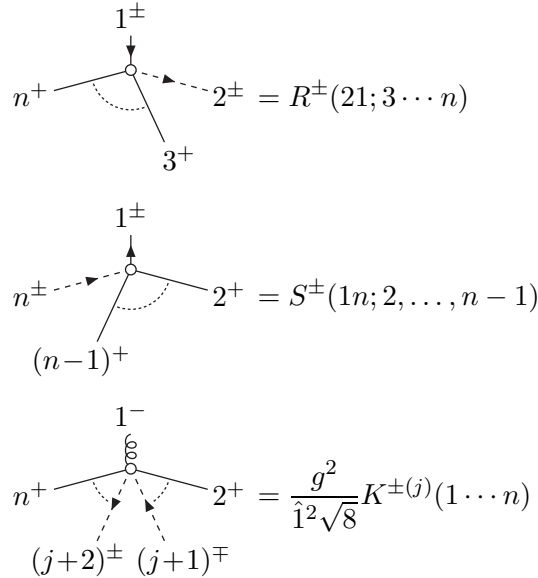

To complete this subsection, we illustrate the new completion vertices
involving fermions arising from this transformation in
fig.~\ref{fig:mhvqcd-completionverts}. These vertices augment the
purely gluonic completion vertices found in ref.\ \cite{Ettle:2007qc}.

\subsection{MHV vertices: an indirect proof}
\label{ssec:offshell-proof}

Let us now return to the discussion begun at the end of section
\ref{ssec:transform} --- specifically, we complete the proof that the
resulting vertices in the transformed lagrangian are MHV vertices,
\ie\ are precisely the MHV amplitudes up to polarisation factors,
continued off shell by the CSW prescription.

We already saw at the end of section \ref{ssec:transform} that on
shell these vertices must be proportional to MHV amplitudes, if we can
ignore the completion vertices. The completion vertices are just the
expansion coefficients $\Upsilon$, $K^\pm$, $R^\pm$ and $S^\pm$,
examples of which are given in fig.~\ref{fig:mhvqcd-completionverts}.
These are indeed involved in the tree level contributions to the
\emph{full} off-shell MHV amplitudes (\ie\ the ones that would then
coincide off shell with the expression from light-cone QCD
\cite{Ettle:2007qc}).  Since by LSZ reduction, the completion vertices
are multiplied by $p^2_1$ (where $p_1$ is the momentum of the field
being expanded), which then vanishes in the on-shell limit, the only
way these contributions can survive the on-shell limit is if the
expansion coefficients themselves diverge in this limit. From their
explicit structure however, it is immediately clear that for
non-vanishing momenta they only have collinear divergences, and thus
for generic values, they make no contribution. (In the kinematically
special case of the three-point vertex with all legs external and on
shell, the momenta are forced to be collinear. The three-point
completion vertices yield the three-point \mhvbar\ amplitude, as
covered in section \ref{ssec:mhvqcd-completion}.)

The remaining question to be settled is whether in the transformed
lagrangian, the vertices can have additional terms that vanish on
shell. We first note, as in ref.\ \cite{Mansfield:2005yd}, that since
the transformation is at equal light-cone time $x^0$ and neither the
remaining vertices of the light-cone lagrangian \eqref{eq:lcqcd-mmp},
\eqref{eq:lcqcd-mmpp} and
\eqref{eq:lcqcd-qbarmq}--\eqref{eq:lcqcd-qbarqqbarq} nor the
transformation, contain $\check\partial$, the resulting vertices must
take the form $\delta(\sum_i {\check p}_i) V$, where $V$ only depends
on the other components of momenta.  It follows that $V$ has no way to
tell whether the on-shell condition $\check p_i = p_i\bar p_i/\hat
p_i$ is satisfied for its momenta. $V$ may however have a term which
on shell reduces to $\sum_i {\check p}_i$ and thus vanishes, \ie\ have
a term of the form $\sum_i p_i\bar p_i/\hat p_i$.  Mansfield proves
for Yang-Mills that the vertices are holomorphic, which trivially
rules out such a term except in the kinematically special case of the
three-point vertex \cite{Mansfield:2005yd}.  The proof could be
readily extended to this case, however it is more direct simply to
note that the remaining vertices of the light-cone lagrangian and the
expansion coefficients are explicitly holomorphic, and thus it is
immediate that the expressions for $V$ are also holomorphic.  Since it
is already clear that the special case of the three-point vertices in
the transformed lagrangian are MHV vertices, this completes the proof
that all the vertices in the transformed lagrangian are MHV vertices.

\subsection{The precise MHV rules/MHV lagrangian correspondence}
\label{ssec:vertex-structure}

Let us now establish the precise relationship between our lagrangian's
vertices and the MHV amplitudes with matching particle and helicity
content. We will need to know this in order to check that the Feynman
rules of the MHV lagrangian we have constructed match the MHV rules
for QCD with massless quarks.

First, let us review the MHV rules for massless quarks in the
fundamental representation, given in ref.\ \cite{Wu:2004jxa} (for the
$U(\NC)$ case). These are as one might expect: the propagator is the
scalar propagator, $i/p^2$, and the off-shell prescription is as in
the purely gluonic case, but one adds two new classes of vertices
(using the expressions for the corresponding amplitudes found in ref.\
\cite{Mangano:1990by}, here adapted to our conventions for
normalisation and fermion operator orderings) to the fold.  The first
class of MHV vertices with massless quarks has either the quark or
antiquark carrying a negative helicity (by conservation of helicity)
and one gluon of negative helicity. Thus they are
\begin{equation}
  \label{eq:csw-ferm-q-gq+}
  \begin{split}
    \begin{matrix}
      \begin{picture}(112,50) \SetOffset(45,5.5) \ArrowLine(30,0)(0,0)
        \ArrowLine(0,0)(-30,0) \Gluon(0,0)(-25.9807,15){2}{5}
        \Gluon(0,0)(0,30){2}{5} \Gluon(0,0)(25.9807,15){2}{5}
        \DashCArc(0,0)(15,35,85){1} \DashCArc(0,0)(15,95,145){1}
        \Vertex(0,0){1.5} \Text(-32,0)[cr]{$1^-$}
        \Text(32,0)[cl]{$n^+$} \Text(0,32)[bc]{$s^-$}
        \Text(27,13)[bl]{$(n\!-\!1)^+$} \Text(-27,13)[br]{$2^+$}
      \end{picture}
    \end{matrix} &= A(1_{{\rm q}}^-,2^+,\dots,
    s^-,\dots,(n-1)^+,n_{\bar{\rm q}}^+) \\
    &= ig^{n-2} \frac{ \langle 1\:s \rangle^3 \langle s\:n
      \rangle}{\langle 1\:2 \rangle \langle 2\:3 \rangle \cdots
      \langle n-1, n\rangle \langle n\:1 \rangle}, \end{split}
\end{equation}
and
\begin{equation}
  \label{eq:csw-ferm-q+gq-}
  \begin{split}
    \begin{matrix}
      \begin{picture}(112,50) \SetOffset(45,5.5) \ArrowLine(30,0)(0,0)
        \ArrowLine(0,0)(-30,0) \Gluon(0,0)(-25.9807,15){2}{5}
        \Gluon(0,0)(0,30){2}{5} \Gluon(0,0)(25.9807,15){2}{5}
        \DashCArc(0,0)(15,35,85){1} \DashCArc(0,0)(15,95,145){1}
        \Vertex(0,0){1.5} \Text(-32,0)[cr]{$1^+$}
        \Text(32,0)[cl]{$n^-$} \Text(0,32)[bc]{$s^-$}
        \Text(27,13)[bl]{$(n\!-\!1)^+$} \Text(-27,13)[br]{$2^+$}
      \end{picture}
    \end{matrix} &= A(1_{{\rm q}}^+,2^+,\dots,
    s^-,\dots,(n-1)^+,n_{\bar{\rm q}}^-) \\
    &= ig^{n-2} \frac{\langle 1\:s \rangle \langle n\:s
      \rangle^3}{\langle 1\:2 \rangle \langle 2\:3 \rangle \cdots
      \langle n-1,n \rangle \langle n\:1 \rangle}. \end{split}
\end{equation}
Treated as amplitudes, these would be associated with the colour
structure
\begin{equation}
  \label{eq:colourstruct-2q}
  (T^{a_{2}} \cdots
  T^{a_{n-1}})_{i_1}{}^{\bar\imath_n}
\end{equation}
in the colour-ordered decomposition.  (We remind the reader that the
full amplitude here would be obtained by summing over permutations of
the gluon labels $2,\dots,n-1$.)  The second class is the MHV vertices
with two quark-antiquark pairs. By conservation of helicity on the
quark lines, these must have all gluons with positive helicity in
order to be of the MHV helicity content. Diagrammatically, these
vertices are
\begin{equation}
  \label{eq:csw-ferm-qgqqgq}
  \begin{split}
    \begin{matrix}
      \begin{picture}(130,84) \SetOffset(54,40)
        \ArrowLine(28.5317,9.2705)(0,0)
        \ArrowLine(0,0)(28.5317,-9.2705)
        \ArrowLine(-28.5317,-9.2705)(0,0)
        \ArrowLine(0,0)(-28.5317,9.2705)
        \Gluon(0,0)(17.6336,24.2705){2}{5}
        \Gluon(0,0)(-17.6336,24.2705){2}{5}
        \Gluon(0,0)(-17.6336,-24.2705){2}{5}
        \Gluon(0,0)(17.6336,-24.2705){2}{5}
        \DashCArc(0,0)(15,59,121){1} \DashCArc(0,0)(15,239,301){1}
        \Vertex(0,0){1.5} \Text(-30,10)[cr]{$1^{h_1}$}
        \Text(30,10)[cl]{$s^{h_s}$} \Text(-30,-10)[cr]{$n^{h_n}$}
        \Text(30,-10)[cl]{$(s\!+\!1)^{h_{s+1}}$}
        \Text(-10,27)[br]{$2^+$} \Text(10,27)[bl]{$(s\!-\!1)^+$}
        \Text(-10,-27)[tr]{$(n\!-\!1)^+$}
        \Text(10,-27)[tl]{$(s\!+\!2)^+$}
      \end{picture}
    \end{matrix}
    &= A(1^{h_1}_{{\rm q}} 2^+ \cdots s^{h_s}_{\bar{\rm q}}
    (s\!+\!1)^{h_{s+1}}_{{\rm q}} \cdots n^{h_n}_{\bar{\rm q}}),
  \end{split}
\end{equation}
and they associate with the colour structure
\begin{equation}
  \label{eq:colourstruct-4q}
  (T^{a_{2}} \cdots
  T^{a_{s-1}})_{i_1}{}^{\bar\imath_s}
  (T^{a_{s+2}} \cdots
  T^{a_{n-1}})_{i_{s+1}}{}^{\bar\imath_n},
\end{equation}
it being understood that when $s=2$ and/or $s=n-2$ that the first
and/or second $T$-string respectively is the identity matrix, and the
partial amplitudes are defined by
\begin{align}
  \label{eq:csw-ferm-4q1}
  \begin{split}
    A(1_{\rm q}^+2^+\cdots s_{\bar {\rm q}}^- s\!+\!1^+_{\rm q}\cdots
    n^-_{\bar {\rm q}}) &= ig^{n-2} \frac {\langle
      s\:n\rangle^2}{\langle1\:2\rangle\langle2\:3\rangle\cdots\langle
      n\:1\rangle}
    \\
    &\quad\times\bigg(\langle 1\: s\rangle\langle
    s\!+\!1,n\rangle+\frac 1 {\NC}\langle s,s\!+\!1\rangle\langle
    n\:1\rangle\bigg),
  \end{split}
  \\
  \label{eq:csw-ferm-4q2}
  \begin{split}
    A(1_{\rm q}^-2^+\cdots s_{\bar {\rm q}}^+ s\!+\!1^-_{\rm q}\cdots
    n^+_{\bar {\rm q}}) &= ig^{n-2} \frac {\langle
      1,s\!+\!1\rangle^2}{\langle1\:2\rangle\langle2\:3\rangle\cdots\langle
      n\:1\rangle}
    \\
    &\quad\times \bigg(\langle 1\: s\rangle\langle
    s\!+\!1,n\rangle+\frac 1 {\NC}\langle s,s+1\rangle\langle
    n\:1\rangle\bigg),
  \end{split}
  \\
  \label{eq:csw-ferm-4q3}
  A(1_{\rm q}^+2^+\cdots s_{\bar {\rm q}}^+ s\!+\!1^-_{\rm q}\cdots
  n^-_{\bar {\rm q}})&=-ig^{n-2} \frac {\langle 1\:s\rangle\langle
    s\!+\!1,n\rangle^3}{\langle1\:2\rangle\langle2\:3\rangle\cdots\langle
    n\:1\rangle}\,,
  \\
  \label{eq:csw-ferm-4q4}
  A(1_{\rm q}^+2^+\cdots s_{\bar {\rm q}}^- s\!+\!1^-_{\rm q}\cdots
  n^+_{\bar {\rm q}}) &= -i\frac{g^{n-2}}{\NC} \frac {\langle s,
    s\!+\!1\rangle^3}{\langle1\:2\rangle\langle2\:3\rangle\cdots
    \langle n\!-\!1,n\rangle}\,.
\end{align}
Note that these are the four independent helicity configurations for
the quarks permitted by helicity conservation along the quark lines
(when the amplitudes are decomposed in this manner, \cf\
\cite{Mangano:1990by} which uses a slightly different arrangement of
the terms).  There are no MHV vertices with more quark-antiquark
lines.

Now, the terms of \eqref{eq:MHVL-massless-q} clearly have colour
structures of the form \eqref{eq:colourstruct-2q} and
\eqref{eq:colourstruct-4q}, and as we noted at the end of section
\ref{ssec:transform}, each MHV amplitude has contributions from only
one vertex in the lagrangian, so at tree-level there is a one-to-one
correspondence between the vertices $V$ of the MHV lagrangian and the
partial amplitudes \eqref{eq:csw-ferm-q-gq+},
\eqref{eq:csw-ferm-q+gq-} and
\eqref{eq:csw-ferm-4q1}--\eqref{eq:csw-ferm-4q4}.  In order to test
this correspondence, we must know the external state polarisation
factors.  The gluon polarisation vectors are given in the spinor
helicity formalism by
\begin{equation}\label{eq:polarization-E}
  E_+ = \sqrt 2 \frac{\nu \tilde\lambda}{\langle \nu\:\lambda \rangle}
  \quad\text{and}\quad
  E_- = \sqrt 2 \frac{\lambda \tilde\nu}{[ \nu\:\lambda ]}
\end{equation}
where $\nu \tilde\nu = \mu$, the null vector normal to the
quantisation surface $\Sigma$. Then in co-ordinates $E_+ = \bar E_-=
-1$ so by the LSZ theorem, when an external $+$ ($-$) polarisation
state is contracted into a $\gA$ ($\bar\gA$) vertex from the
lagrangian, it contributes a factor of
\begin{equation}
  -1 \times -\frac{ig}{\sqrt 2},
\end{equation}
where the second factor restores the canonical normalisation of the
gauge field from \eqref{eq:gauge-conv}. This means that the
correspondence between the vertex $V^s_{\text{YM}}(1 \cdots n)$ and
the pure-gluon partial amplitude $A(1^-2^+\cdots s^- \cdots n^+)$
proceeds according to:
\begin{align*}
  \frac{4i}{g^2} \times \left( -1 \times -\frac{ig}{\sqrt 2} \right)^n
  \times V^s_{\text{YM}}(1 \cdots n) &= ig^{n-2} \frac{\langle 1\:s
    \rangle^4}{\langle 1\:2 \rangle \langle 2\:3 \rangle \cdots
    \langle
    n\:1 \rangle} \\
  &= A(1^-2^+\cdots s^- \cdots n^+),
\end{align*}
that is,
\[
V^s_{\text{YM}}(1 \cdots n) = (-i\sqrt 2)^{n-4} \frac{\langle 1\:s
  \rangle^4}{\langle 1\:2 \rangle \langle 2\:3 \rangle \cdots \langle
  n\:1 \rangle},
\]
which is the expression for $V^s_{\text{YM}}(1 \cdots n)$ given in
\cite{Ettle:2006bw}\footnote{Adjusted for the normalisation in use
  here.}. (Of course, this follows immediately following the
discussion below \eqref{eq:Abar0-series}.)

For the quark correspondence, let us consider first the LSZ reduction
\emph{before} we remove the non-dynamical fermionic degrees of
freedom. For example, in this context, an out-going $+$ helicity quark
with momentum $p$ produces a term
\begin{equation}
  \label{eq:decouple-massless}
  \bar\varphi(p)_{\dot\alpha} (-i p\cdot\sigma)^{\dot\alpha \beta}
  \langle \cdots \begin{pmatrix} \beta^- \\ \alpha^- \end{pmatrix}_\beta \cdots \rangle,
\end{equation}
where $\bar\varphi(p)$ is the appropriate polarisation spinor of
\eqref{eq:poln-qp}, and $-i p\cdot\sigma$ is the inverse of the
propagator obtained from \eqref{eq:lcqcd}. We can now compute the
correlation function using the partition function furnished by the MHV
lagrangian. Since $\varphi_1 \equiv \beta^-$ is replaced by its
equation of motion, one might expect this non-propagating component to
complicate things. Thankfully
\begin{equation}
  \label{eq:invprop}
  \bar\varphi(p) (-i p\cdot\sigma) = - \frac i{2^{1/4} \sqrt{\hat p}}
  (0, p^2),
\end{equation}
so it does not arise in the computation. We proceed to replace
$\varphi_2 \equiv \alpha^-$ with its expression in terms of the new
variables, and note that momentum conservation implies only the
leading order term $\xi^-$ survives the on-shell limit for generic
momenta (\ie, we note that the $S$-matrix equivalence theorem applies
here).  The propagator $\langle \xi^- \bar\xi^+ \rangle$ of
\eqref{eq:propagators} cancels factors in \eqref{eq:invprop} to leave
a polarisation factor
\begin{equation}\label{eq:polarization-fermion}
  2^{1/4}\sqrt{\hat p}.
\end{equation}
One may show similarly that the same expression applies for the $-$
helicity state, and for the antiquarks. In summary, we state the
polarisation spinors and the fields in the lagrangian associated with
each out-going state in table \ref{tbl:field-state}. We note here that
this accounts for the use of a \emph{scalar} propagator in the MHV
rules for fermions, rather than the fermion propagators we gave in
\eqref{eq:propagators} which joins the vertices of the MHV
lagrangian. In the former the numerator $\sqrt 2 \hat p$ of
\eqref{eq:propagators} has been absorbed into the vertices at either
end as the polarisation factors to yield the familiar MHV amplitudes.

\TABLE[h]{%
  \hspace{2cm} 
  \begin{tabular}{lccc}
    \toprule
    \multicolumn{2}{l}{\textbf{State}} &
    \textbf{Polarisation} & \textbf{Field} \\
    \midrule
    \multirow{2}{*}{particle}
    & $+$ & $\bar\varphi(p)$ & $\bar\xi^+$ \\
    & $-$ & $\omega(p)$      & $\bar\xi^-$ \\
    \multirow{2}{*}{antiparticle}
    & $+$ & $\bar\omega(p)$  & $\xi^+$ \\
    & $-$ & $\varphi(p)$     & $\xi^-$ \\
    \bottomrule
  \end{tabular}
  \hspace{2cm} 
  \label{tbl:field-state}
  \caption{The polarisation spinor and lagrangian field associated
    with each out-going quark state.}
}

Now consider the MHV amplitude with one quark-antiquark pair. We will
choose its external state to be
\[
\langle 0 \rvert q^\pm_1 A^+_2 \cdots A^-_s \cdots A^+_{n-1} \bar
q^\mp_n,
\]
where $\langle 0 \rvert$ is the (asymptotic) free vacuum.  This
contracts into one of the vertices in the first term of
\eqref{eq:MHVL-massless-q} (depending on the quarks' helicities), and
multiplies it by an external state factor of
\[
(-1) \times \frac{4i}{g^2} \times \left(\frac{ig}{\sqrt2}\right)^{n-2}
\times 2^{1/4}\sqrt{\hat1} \times 2^{1/4}\sqrt{\hat n}.
\]
Considering the factors delimited by $\times$ symbols, the first comes
from Fermi statistics; the second from the path integral; the third
from gluon polarisation and normalisation; and the final two from the
external state spinors (see above). Thus we have
\begin{equation}
  \label{eq:A-V-2q}
  V^{s,\pm\mp}_\text{F}(1\cdots n) =
  \frac{2^{(n-7)/2}g^{4-n}}{i^{n+1} \sqrt{\hat 1} \sqrt{\hat n}}
  A(1_{{\rm q}}^\pm,2^+,\dots, s^-, \dots,(n\!-\!1)^+,n_{\bar{\rm
      q}}^\mp);
\end{equation}
concretely,
\begin{align}
  \label{eq:A-V-+-}
  V^{s,+-}_\text{F}(1\cdots n) &= \frac{g^2}{i^n\sqrt 8} \frac{\hat 2
    \cdots \widehat{n-1}\:
    (1\:s)(n\:s)^3}{\hat s^2 \hat n \: (1\:2) \cdots (n-1,n)(n\:1)},\\
  \label{eq:A-V--+}
  V^{s,-+}_\text{F}(1\cdots n) &= \frac{g^2}{i^n\sqrt 8} \frac{\hat 2
    \cdots \widehat{n-1}\: (1\:s)^3(s\:n)}{\hat1 \hat s^2 \: (1\:2)
    \cdots (n-1,n)(n\:1)},
\end{align}
using \eqref{eq:lc-spinorbrackets}.  For vertices with two
quark-antiquark pairs, we choose the external state
\[
\langle 0 \rvert q^{h_1}_1 A^+_2 \cdots A^+_{s-1} \bar q^{h_s}_{s} \:
q^{h_{s+1}}_{s+1} A_{s+2}^+ \cdots A^+_{n-1} \bar q^{h_n}_n.
\]
By contracting into the second term of \eqref{eq:MHVL-massless-q} as
appropriate to the helicity content, we arrive at
\begin{equation}
  \label{eq:A-V-4q}
  V_\text{F}^{s,h_1h_sh_{s+1}h_n}(1 \cdots n) = \frac{2^{n/2-5}
    g^{6-n}}{i^{n+1} \sqrt{\hat1} \sqrt{\hat s} \sqrt{\widehat{s+1}}
    \sqrt{\hat n}} A(1^{h_1}_{{\rm q}} 2^+ \cdots s^{h_s}_{\bar{\rm q}}
  (s\!+\!1)^{h_{s+1}}_{{\rm q}} \cdots n^{h_n}_{\bar{\rm q}}),
\end{equation}
where the $A$ on the RHS are given by
\eqref{eq:csw-ferm-4q1}--\eqref{eq:csw-ferm-4q4}.  Note that our
arrangement of the colour structure has combined the terms such that
$V_\text{F}^{s,+-+-}$ and $V_\text{F}^{s,-+-+}$ have both ${\cal
  O}(1)$ and ${\cal O}(1/\NC)$ pieces, whereas $V_\text{F}^{s,++--}$
has only a ${\cal O}(1)$ piece and $V_\text{F}^{s,+--+}$ has only a
${\cal O}(1/\NC)$ piece.

\section{Example Vertices}
\label{sec:examples}

In this section, we will give some explicit examples of the terms in
our MHV lagrangian, and verify explicitly that these terms are
proportional off shell to the known tree-level MHV partial amplitudes
up to external polarisation factors (\ie\ that they satisfy
\eqref{eq:A-V-+-}--\eqref{eq:A-V-4q}).

\subsection{Two quarks and two gluons}

Let us begin by considering the vertex $V^{3,+-}_\text{F}(1234)$ that
couples two gluons of opposing helicities to a quark-antiquark pair,
\ie\ the sole contributor to $A(1_{\rm q}^+ 2^+ 3^- 4_{\bar{\rm
    q}}^-)$.  Looking at table \ref{tbl:mhvqcd-content}, we see that,
based upon field content, the term we are considering receives
contributions from $L^{\mmp}$, $L^{\bar\psi+-\psi}$ and
$L^{\bar\psi-\psi}$ (in \eqref{eq:lcqcd-mmp}, \eqref{eq:lcqcd-qbarpmq}
and \eqref{eq:lcqcd-qbarmq} respectively). Let us consider each in
turn.  Written in momentum space, \eqref{eq:lcqcd-mmp} is
\begin{equation}
  \label{eq:lcqcd-mmp-M}
  L^{\mmp} = i \tr \int_{123} \frac{\hat 3}{\hat 1 \hat 2} (1\:2) \:
  \bar\gA_{\bar 1} \bar\gA_{\bar 2} \gA_{\bar 3},
\end{equation}
where here and in the foregoing, a momentum-conserving $\delta$
function of the sum of all the momenta in the integral measure is
omitted for clarity.  The term with the relevant colour structure is
extracted by substituting for the second $\bar\gA$ with the
lowest-order $\bar\xi^+\xi^-$ term in
\eqref{eq:AbarF-series}. Carefully relabelling, this term is
\begin{equation}
  \label{eq:qqgg-1}
  - \frac{g^2}{\sqrt 8} \int_{1234} \frac{\hat 2 \hat 4}{\hat 3 (\hat 1 +
    \hat 4)^2} \frac{(3\:2)}{(4\:1)} \: \bar\xi^+_{\bar 1} \gB_{\bar 2} \bar\gB_{\bar 3} \xi^-_{\bar 4}.
\end{equation}
Next, consider the contribution from $L^{\bar\psi+-\psi}$: here, only
the leading order substitution are needed for the fields involved so
we simply extract the relevant term from the momentum-space
representation of \eqref{eq:lcqcd-qbarpmq}, giving
\begin{equation}
  \label{eq:qqgg-2}
  \frac{g^2}{\sqrt 8} \int_{1234} \frac{\hat 3 - \hat 2}{(\hat 1 + \hat
    4)^2} \: \bar\xi^+_{\bar 1} \gB_{\bar 2} \bar\gB_{\bar 3} \xi^-_{\bar 4}.
\end{equation}
Finally $L^{\bar\psi-\psi}$, which in momentum space is
\begin{equation}
  \label{eq:lcqcd-qbarmq-M}
  L^{\bar\psi-\psi} = -\frac{ig^2}{\sqrt 8} \int_{123} \left\{ \left(
      \frac{\tilde 3}{\hat 3} - \frac{\tilde 2}{\hat 2} \right)
    \bar\alpha^+_{\bar 1} \bar\gA_{\bar 2} \alpha^-_{\bar 3} + 
    \left(
      \frac{\tilde2 + \tilde3}{\hat2 + \hat3} - \frac{\tilde 2}{\hat 2} \right)
    \bar\alpha^-_{\bar 1} \bar\gA_{\bar 2} \alpha^+_{\bar 3}
  \right\},
\end{equation}
contributes two terms from the next-to-leading order substitutions for
$\bar\alpha^+$ and $\bar\gA$ from \eqref{eq:fqold-in-new} and
\eqref{eq:Abar0-series}, leading to
\begin{equation}
  \frac{g^2}{\sqrt 8} \int_{1234} \left\{
    \frac{\hat1 + \hat2}{\hat3 \hat4} \frac{(4\:3)}{(1\:2)}
    + \frac{\hat3^2}{\hat4 (\hat1 + \hat4)^2} \frac{(1\:4)}{(2\:3)}
  \right\} \bar\xi^+_{\bar 1} \gB_{\bar 2} \bar\gB_{\bar 3} \xi^-_{\bar 4}.
\end{equation}
Summing over the coefficients, accounting for conservation of momentum
and the normalization of $\cal B$, and comparing with
\eqref{eq:MHVL-massless-q}, one can read off the MHV vertex:
\begin{equation}
  V^{3,+-}_\text{F}(1234)=-\frac{g^2}{\sqrt 8} \frac{\hat 2}{\hat3 \hat4}
  \frac{(1\:3) (4\:3)^2}{(1\:2) (2\:3) (4\:1)},
\end{equation}
and it is immediate to see that this satisfies \eqref{eq:A-V-2q}.

One may treat the remaining vertices similarly, by considering the
other possible choices of substitution for $\bar\gA$.  We reproduce
these vertices below:
\begin{align*}
  V^{3,-+}_\text{F}(1234) &= \hphantom{-} \frac{g^2}{\sqrt 8}
  \frac{\hat 2}{\hat1 \hat3} \frac{(1\:3)^3} {(1\:2)(2\:3)(4\:1)},
  \\
  V^{2,+-}_\text{F}(1234) &= - \frac{g^2}{\sqrt 8} \frac{\hat3}{\hat2
    \hat4} \frac{(2\:4)^3} {(2\:3)(3\:4)(4\:1)},
  \\
  V^{2,-+}_\text{F}(1234) &= \hphantom{-} \frac{g^2}{\sqrt 8}
  \frac{\hat3}{\hat1 \hat2} \frac{(1\:2)^2 (2\:4)}
  {(2\:3)(3\:4)(4\:1)}.
\end{align*}
It is easy to see that these satisfy \eqref{eq:A-V-+-} and
\eqref{eq:A-V--+}.

\subsection{Four quarks}

\label{ssec:4q}
In amplitudes with two or more quark-antiquark pairs, terms ${\cal
  O}(1/N_{\rm C})$ and higher contribute at the tree-level through the
sub-leading colour structures, and we must retain these terms
throughout the analysis. These terms are generated by the ${\rm
  SU}(N_{\rm C})$ Fierz identity
\[ (T^a)_i^{\bar\jmath} (T^a)_k^{\bar l} = \delta_i^{\bar l}
\delta_k^{\bar\jmath} - \frac1{N_{\rm C}} \delta_i^{\bar\jmath}
\delta_k^{\bar l},
\]
where second term on the RHS comes about because the generators of
${\rm SU}(N_{\rm C})$ are traceless\footnote{This is equivalent to
  subtracting the `fictitious photon', as found in the literature that
  works with a ${\rm U}(N_{\rm C})$ gauge symmetry instead}.  We note
that the four quark vertices $V^{s,h_1h_2h_3h_4}_\text{F}(1234)$
receive contributions from $L^{\bar\psi-\psi}$ and
$L^{\bar\psi\psi\bar\psi\psi}$. In momentum space, these are given by
\eqref{eq:lcqcd-qbarmq-M} and
\begin{equation}
  \label{eq:lcqcd-qbarqqbarq-M}
  \begin{split}
    L^{\bar\psi\psi\bar\psi\psi} = \frac{g^4}8 \int_{1234} \Biggl\{ &
    \left( \frac1{(\hat1 + \hat4)^2} + \frac 1{N_{\rm C}}
      \frac1{(\hat1 + \hat2)^2} \right) (\bar\alpha^-_{\bar1}
    \alpha^+_{\bar2} \bar\alpha^-_{\bar3} \alpha^+_{\bar4} +
    \bar\alpha^+_{\bar1} \alpha^-_{\bar2} \bar\alpha^+_{\bar3}
    \alpha^-_{\bar4}) \\
    &+ \frac2{(\hat1 + \hat4)^2} \bar\alpha^+_{\bar1} \alpha^+_{\bar2}
    \bar\alpha^-_{\bar3} \alpha^-_{\bar4} + \frac1{N_{\rm C}}
    \frac1{(\hat1 + \hat2)^2} \bar\alpha^-_{\bar1} \alpha^+_{\bar2}
    \bar\alpha^+_{\bar3} \alpha^-_{\bar4} \Biggr\}.
  \end{split}
\end{equation}
We substitute for $\bar\gA$ in \eqref{eq:lcqcd-qbarmq-M} using the
leading order terms in \eqref{eq:AbarF-series} and for the fermions in
\eqref{eq:lcqcd-qbarqqbarq-M}. Summing and symmetrising over the
momentum labels as much as possible leads to the following terms in
the MHV lagrangian:
\begin{equation}
  \label{eq:lmhv-qqqq-M}
  \begin{split}
    \frac{g^4}8 \int_{1234} \Biggl\{ & \frac12 \left(
      \frac{(2\:4)^2}{\hat2 \hat4 (1\:4) (3\:2)} + \frac 1{N_{\rm C}}
      \frac{(2\:4)^2}{\hat2 \hat4 (1\:2) (3\:4)} \right) \:
    \bar\xi^+_{\bar1}{}^{\bar\imath_1} \xi^-_{\bar2}{}_{i_2}
    \bar\xi^+_{\bar3}{}^{\bar\imath_3} \xi^-_{\bar4}{}_{i_4} \\ & +
    \frac12 \left( \frac{(1\:3)^2}{\hat1 \hat3 (1\:4) (3\:2)} + \frac
      1{N_{\rm C}} \frac{(1\:3)^2}{\hat1 \hat3 (1\:2) (3\:4)} \right)
    \: \bar\xi^-_{\bar1}{}^{\bar\imath_1} \xi^+_{\bar2}{}_{i_2}
    \bar\xi^-_{\bar3}{}^{\bar\imath_3} \xi^+_{\bar4}{}_{i_4} \\ &
    -\frac{(3\:4)^2}{\hat3 \hat4 (1\:4) (3\:2)} \:
    \bar\xi^+_{\bar1}{}^{\bar\imath_1} \xi^+_{\bar2}{}_{i_2}
    \bar\xi^-_{\bar3}{}^{\bar\imath_3} \xi^-_{\bar4}{}_{i_4} -
    \frac1\NC \frac{(2\:3)^2}{\hat2 \hat3 (1\:2)(3\:4)}
    \bar\xi^+_{\bar1}{}^{\bar\imath_1} \xi^-_{\bar2}{}_{i_2}
    \bar\xi^-_{\bar3}{}^{\bar\imath_3} \xi^+_{\bar4}{}_{i_4} \Biggr\}
    \di12 \di34.
  \end{split}
\end{equation}
Writing it out this way makes immediate contact with the second term
of \eqref{eq:MHVL-massless-q}, and we can read off the vertices
immediately:
\begin{align}
  V^{s,+-+-}_\text{F}(1234) &= \frac{g^4}8 \left(
    \frac{(2\:4)^2}{\hat2 \hat4 (1\:4) (3\:2)} + \frac 1{N_{\rm C}}
    \frac{(2\:4)^2}{\hat2 \hat4 (1\:2) (3\:4)} \right),
  \\
  V^{s,-+-+}_\text{F}(1234) &= \frac{g^4}8 \left(
    \frac{(1\:3)^2}{\hat1 \hat3 (1\:4) (3\:2)} + \frac 1{N_{\rm C}}
    \frac{(1\:3)^2}{\hat1 \hat3 (1\:2) (3\:4)} \right),
  \\
  V^{s,++--}_\text{F}(1234) &= -\frac{g^4}8 \frac{(3\:4)^2}{\hat3
    \hat4 (1\:4) (3\:2)},
  \\
  V^{s,+--+}_\text{F}(1234) &= -\frac{g^4}8 \frac1\NC
  \frac{(2\:3)^2}{\hat2 \hat3 (1\:2)(3\:4)},
\end{align}
which are readily seen to concur with \eqref{eq:A-V-4q} upon using
\eqref{eq:csw-ferm-4q1}--\eqref{eq:csw-ferm-4q4} and
\eqref{eq:lc-spinorbrackets}.

\subsection{Two quarks and three gluons}

Finally, we consider the vertex $V^{3,+-}_\text{F}(12345)$.  In the
MHV lagrangian, this is the coefficient of the term in
$\bar\xi^+_{\bar 1} \gB_{\bar 2} \bar\gB_{\bar 3} \gB_{\bar 4}
\xi^-_{\bar 5}$ and it receives contributions from $L^\mmp$,
$L^{--++}$, $L^{\bar\psi-\psi}$ and $L^{\bar\psi+-\psi}$, so we will
write it as
\begin{equation}
  \label{eq:qgggq-struct}
  \int_{12345} (W^\mmp + W^{--++} + W^{\bar\psi-\psi} + W^{\bar\psi+-\psi})
  \bar\xi^+_{\bar 1} \gB_{\bar 2}
  \bar\gB_{\bar 3} \gB_{\bar 4} \xi^-_{\bar 5}.
\end{equation}
Let us consider each of the $W$s in turn. First, we observe from
\eqref{eq:lcqcd-mmp-M} and the structures of \eqref{eq:Abar0-series}
and \eqref{eq:AbarF-series} that $L^\mmp$ yields four terms with the
structure of \eqref{eq:qgggq-struct} coming from the different
possible choices substitution for $\gA$. We carry this out and
carefully relabel the momenta while accounting for the anticommuting
nature of the fermions to obtain
\newcommand\gintl[1]{\frac{#1g^2}{\sqrt 8}\int_{12345} \Biggl\{}
\begin{multline}
  \label{eq:qgggq-1}
  W^\mmp = \frac{ig^2}{\sqrt 8} \Biggl\{ \frac{\hat2 \hat3^2 \hat5
    (2,3+4)}{(\hat1 + \hat5)^3 (\hat3 + \hat4)^3}
  \Upsilon(-,3,4) \Upsilon(-,5,1) \\
  + \frac{\hat3^2 \hat4 \hat5 (2+3,4)}{(\hat1 + \hat5)^3 (\hat2 +
    \hat3)^3}
  \Upsilon(-,2,3) \Upsilon(-,5,1) \\
  - \frac{\hat4 \hat5 (3\:4)}{\hat3 (\hat3 + \hat4)^3}
  \Upsilon(-,5,1,2) + \frac{\hat2 \hat5 (2\:3)}{\hat3 (\hat2 +
    \hat3)^3} \Upsilon(-,4,5,1) \Biggr\}.
\end{multline}
We remind the reader here that in the argument lists of $\Upsilon$,
$K$, \etc, $-$ is a placeholder whose value should be taken to be the
negative of the sum of the other momenta passed to that coefficient.

Next, $L^{--++}$ has terms of the form $\tr (\bar\gA \gA \bar\gA \gA)$
and $\tr (\bar\gA \bar\gA \gA \gA)$, of which only the former
contribute terms in \eqref{eq:qgggq-struct}. In momentum space, this
is
\begin{equation}
  \tr \int_{1234} \left\{ \frac{\hat2 \hat3}{(\hat3 + \hat4)^2}
    + \frac{\hat3 \hat4}{(\hat2 + \hat3)^2} \right\} \bar\gA_{\bar 1}
  \gA_{\bar 2} \bar\gA_{\bar 3} \gA_{\bar 4}.
\end{equation}
Substituting for each $\bar\gA$ in turn using the lowest-order terms
in \eqref{eq:AbarF-series} gives
\begin{multline}
  \label{eq:qgggq-2}
  W^{--++} = \frac{g^2}{\sqrt 8} \frac{\hat5}{(\hat1 + \hat5)^2}
  \Biggl\{ \left( \frac{\hat2 \hat3}{(\hat3 + \hat4)^2} + \frac{\hat3
      \hat4}{(\hat2 + \hat3)^2} \right) \Upsilon(-,5,1) \\
  + \left( \frac{\hat4 (\hat1 + \hat5)}{(\hat3 + \hat4)^2} +
    \frac{\hat2 (\hat1 + \hat5)}{(\hat2 + \hat3)^2} \right)
  \Upsilon(-,5,1) \Biggr\}.
\end{multline}

$L^{\bar\psi-\psi}$ contributes four terms in the structure of
\eqref{eq:qgggq-struct}, owing to the fact that we will now also see
terms from the fermion series expansions \eqref{eq:fqold-in-new} and
\eqref{eq:fpold-in-new} when these are substituted into
\eqref{eq:lcqcd-qbarmq-M}. Upon re-arrangement and re-labelling of the
momenta, we arrive at the contribution
\begin{multline}
  \label{eq:qgggq-3}
  W^{\bar\psi-\psi} = -\frac{ig^2}{\sqrt 8} \Biggl\{ \left(
    \frac{\tilde4 + \tilde5}{\hat4 + \hat5} - \frac{\tilde3}{\hat3}
  \right) \frac{\hat5}{\hat4 + \hat5} \Upsilon(-,1,2) \Upsilon(-,4,5) \\
  + \left( \frac{\hat3}{\hat3 + \hat4} \right)^2 \left(
    \frac{\tilde5}{\hat5} - \frac{\tilde3 + \tilde4}{\hat3 + \hat4}
  \right) \Upsilon(-,1,2)  \Upsilon(-,3,4) \\
  + \left( \frac{\hat3}{\hat2 + \hat3 + \hat4} \right)^2 \left(
    \frac{\tilde5}{\hat5} - \frac{\tilde1 + \tilde5}{\hat1 + \hat5}
  \right) \Upsilon(-,2,3,4)
  \\
  + \left( \frac{\hat3}{\hat2 + \hat3} \right)^2 \left( \frac{\tilde4
      + \tilde5}{\hat4 + \hat5} - \frac{\tilde2 + \tilde3}{\hat2 +
      \hat3} \right) \frac{\hat5}{\hat4 + \hat5} \Upsilon(-,2,3)
  \Upsilon(-,4,5) \Biggr\}.
\end{multline}

Finally, $L^{\bar\psi+-\psi}$, which has momentum-space representation
\begin{multline}
  \label{eq:lcqcd-qbarpmq-M}
  L^{\bar\psi+-\psi} = -\frac{g^2}{\sqrt 8} \int_{1234} \Biggl\{
  \left( \frac 1{\hat3 + \hat4} + \frac{\hat2 - \hat3}{(\hat2 +
      \hat3)^2} \right) \bar\alpha^+_{\bar1} \bar\gA_{\bar2}
  \gA_{\bar3}
  \alpha^-_{\bar4} \\
  + \frac{\hat2 - \hat3}{(\hat2 + \hat3)^2} \bar\alpha^+_{\bar1}
  \gA_{\bar2} \bar\gA_{\bar3} \alpha^-_{\bar4} + \text{l.h.\ pieces}
  \Biggr\},
\end{multline}
contributes four terms to \eqref{eq:qgggq-struct} from substitutions
for $\bar\gA$ and the fermions:
\begin{multline}
  \label{eq:qgggq-4}
  W^{\bar\psi+-\psi} = - \frac{g^2}{\sqrt 8} \Biggl\{ \left( \frac
    1{\hat4 + \hat5} + \frac{\hat3 - \hat4}{(\hat3 + \hat4)^2} \right)
  \Upsilon(-,1,2) + \frac{\hat2 - \hat3}{(\hat2 + \hat3)^2}
  \frac{\hat5}{\hat4 + \hat5} \Upsilon(-,4,5) \\
  + \left( \frac{\hat3}{\hat3 + \hat4} \right)^2 \frac{\hat2 -
    \hat3 - \hat4}{(\hat2 + \hat3 + \hat4)^2}  \Upsilon(-,3,4) \\
  +\left( \frac{\hat3}{\hat2 + \hat3} \right)^2 \left( \frac 1{\hat4 +
      \hat5} + \frac{\hat2 + \hat3 - \hat4}{(\hat2 + \hat3 + \hat4)^2}
  \right) \Upsilon(-,2,3) \Biggr\}.
\end{multline}
$V^{+-,3}_\text{F}(12345)$ is obtained from the sum of the
coefficients in \eqref{eq:qgggq-1}, \eqref{eq:qgggq-2},
\eqref{eq:qgggq-3} and \eqref{eq:qgggq-4} and is
\begin{equation}
  V^{+-,3}_\text{F}(12345) = 
  \frac{ig^2}{\sqrt 8} \frac{\hat2 \hat4}{\hat3 \hat5}
  \frac{(1\:3) (3\:5)^3}{(1\:2) (2\:3) (3\:4) (4\:5) (5\:1)},
\end{equation}
which satisfies \eqref{eq:A-V-2q}, so the vertex is proportional to
the MHV amplitude.

\subsection{On completion vertices and missing amplitudes}
\label{ssec:mhvqcd-completion}

We showed in \cite{Ettle:2007qc} that the $S$-matrix receives
contributions from more than just the vertex content of a MHV
lagrangian.  The complete treatment of the $S$-matrix via the LSZ
reduction revealed completion vertices that arose from the terms in
the transformation itself.  We found that they are required to obtain
amplitudes that can only be formed by using vertices eliminated by the
field transformation.  At tree-level, these can be non-vanishing on
shell in $(2,2)$ signature or for complex momenta and they are needed
for the complete recovery of the full off-shell theory at tree-level.
At loop level, they are needed in general to recover the full physical
on-shell amplitudes.  We demonstrated this by calculating the
$A(1^-2^+3^+)$ tree-level amplitude and showing that $A(1^+2^+3^+4^+)$
at one-loop matches the known result, as obtained using light-cone
Yang--Mills theory.

\begin{figure}[t]
  \centering \subfigure{
    \begin{picture}(100,80) \SetOffset(50,27) \Gluon(0,27)(0,40){2}{2}
      \ArrowLine(34.6410,-20)(23.3827,-13.5)
      \ArrowLine(-23.3827,-13.5)(-34.6410,-20)
      \DashArrowLine(21.6506,-12.5)(0,25){1}
      \DashArrowLine(0,25)(-21.6506,-12.5){1} \BCirc(0,25){2}
      \BCirc(21.6506,-12.5){2} \BCirc(-21.6506,-12.5){2}
      \Text(0,43)[bc]{$2^-$} \Text(35.6410,-19)[tl]{$3_{{\rm q}}^+$}
      \Text(-35.6410,-19)[tr]{$1_{\bar{\rm q}}^-$}
      \Text(-4,24)[tr]{$+$} \Text(4,24)[tl]{$-$}
      \Text(-19.0,-8.5)[br]{$-$} \Text(19.0,-8.5)[bl]{$+$}
    \end{picture}
    \label{fig:mhv-completion-qgq-1}
  }\quad \subfigure{
    \begin{picture}(100,80) \SetOffset(50,27) \Gluon(0,27)(0,40){2}{2}
      \ArrowLine(34.6410,-20)(23.3827,-13.5)
      \ArrowLine(-23.3827,-13.5)(-34.6410,-20)
      \Line(21.6506,-12.5)(0,25)
      \DashArrowLine(21.6506,-12.5)(-21.6506,-12.5){1} \BCirc(0,25){2}
      \BCirc(21.6506,-12.5){2} \BCirc(-21.6506,-12.5){2}
      \Text(4,24)[tl]{$-$} \Text(19.0,-8.5)[bl]{$+$}
      \Text(-19,-13.5)[tl]{$-$} \Text(19,-13.5)[tr]{$+$}
    \end{picture}
    \label{fig:mhv-completion-qgq-2}
  }\quad \subfigure{
    \begin{picture}(100,100) \SetOffset(50,27)
      \Gluon(0,27)(0,40){2}{2} \ArrowLine(34.6410,-20)(23.3827,-13.5)
      \ArrowLine(-23.3827,-13.5)(-34.6410,-20)
      \Line(0,25)(-21.6506,-12.5)
      \DashArrowLine(21.6506,-12.5)(-21.6506,-12.5){1} \BCirc(0,25){2}
      \BCirc(21.6506,-12.5){2} \BCirc(-21.6506,-12.5){2}
      \Text(-4,24)[tr]{$-$} \Text(-19.0,-8.5)[br]{$+$}
      \Text(-19,-13.5)[tl]{$-$} \Text(19,-13.5)[tr]{$+$}
    \end{picture}
    \label{fig:mhv-completion-qgq-3}
  }
  \caption{Contributions to the tree-level $A(1_{{\rm q}}^+ 2^+
    3_{\bar{\rm q}}^-)$ amplitude, before applying LSZ reduction. All
    momenta are directed out of the diagrams, arrows indicate colour
    flow.}
  \label{fig:mhvqcd-completion-qgq}
\end{figure}
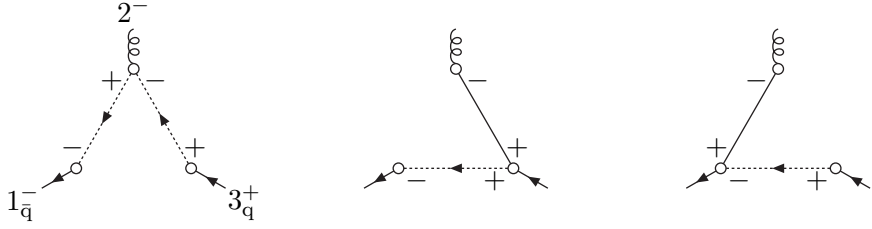

The situation when one adds quarks is no different: quark-gluon
amplitudes whose construction requires erstwhile \mhvbar-like vertices
are recovered through completion vertices found in
\eqref{eq:A-series}, \eqref{eq:fpold-in-new}, \eqref{eq:fqold-in-new}
and \eqref{eq:AbarF-series}. As an example of this, let us study the
partial amplitude $A(1_{{\rm q}}^+ 2^+ 3_{\bar{\rm q}}^-)$. It is
given by
\begin{equation}
  A(1_{{\rm q}}^+ 2^+ 3_{\bar{\rm q}}^-) = ig \frac{[2\:1]^2}{[3\:1]}.
  \label{eq:qcd-completion}
\end{equation}
The LSZ reduction gives this amplitude as
\begin{equation}
  \label{eq:qcd-completion-1}
  A(1_{{\rm q}}^+ 2^+ 3_{\bar{\rm q}}^-) =\lim_{p_1^2, p_2^2, p_3^2
    \rightarrow 0}
  ip_2^2 \times \frac{-ip_1^2}{2^{1/4}\sqrt1} \times
  \frac{-ip_3^2}{2^{1/4}\sqrt3} \times \langle \alpha^-_1 \, \bar A_2 \,
  \bar\alpha^+_3 \rangle.
\end{equation}
Here, the first factor contributes the gluon polarisation and inverse
propagator; and the second and third factors are \eqref{eq:invprop}
for the quarks.  The correlation function may be computed by
substituting for each of the fields involved with their
next-to-leading-order expressions from \eqref{eq:fpold-in-new},
\eqref{eq:fqold-in-new} and \eqref{eq:AbarF-series}, or equivalently
by evaluating the sum of the three diagrams of
fig.~\ref{fig:mhvqcd-completion-qgq}, constructed from the vertices of
fig.~\ref{fig:mhvqcd-completionverts}.
Thus, \eqref{eq:qcd-completion-1} becomes
\begin{equation}\begin{split}
    A(1_{{\rm q}}^+ 2^+ 3_{\bar{\rm q}}^-) &= -\frac{i}{\sqrt{2 \:
        \hat1 \hat3}} p_1^2 p_2^2 p_3^2 \Biggl\{ \frac{\hat1}{p_1^2}
    \frac{\hat3}{p_3^2} \frac{ig}{\hat2^2} \hat3 \Upsilon(231) +
    \frac{\hat1}{p_1^2} \frac1{p_2^2} \: ig \: \Upsilon(312) +
    \frac1{p_2^2} \frac{\hat3}{p_3^2} \: ig \:
    \frac{\hat3}{\hat1} \Upsilon(123) \Biggr\} \\
    &= \frac{ig}{\sqrt2} \sqrt{\hat 1 \hat 3} \frac{\hat3}{(1\:2)}
    \left( \frac{p_1^2}{\hat 1} + \frac{p_2^2}{\hat 2} +
      \frac{p_3^2}{\hat 3} \right) \\
    &= ig \sqrt2 \frac1{\hat2} \sqrt{\frac{\hat3}{\hat1}} \: \{3\:1\}.
  \end{split}\end{equation}
where we have used \eqref{eq:Rr-coeff}, \eqref{eq:Sr-coeff},
\eqref{eq:Sl-coeff}, \eqref{eq:K--coeff} and \eqref{eq:K+-coeff}, and
restored the canonical normalisation of $A$ using
\eqref{eq:gauge-conv}.  This expression may be shown to equal
\eqref{eq:qcd-completion} using \eqref{eq:lc-spinorbrackets}. We
obtained the final line using the result \cite{Ettle:2007qc}
\begin{equation}
  \label{eq:sum-bilinears}
  \sum_j\frac{(p\:j)\mdot\{j\:q\}}{\hat j} =
  \frac{\hat p\,\hat q}{2} \sum_j \frac{p^2_j}{\hat j}
\end{equation}
which holds for any set $\{p_i\}$ of momenta that sums to zero.

\section{Massive Quarks}
\label{sec:masses}

In this section, we analyse the MHV QCD lagrangian in the case where
the quarks have masses. For simplicity, we will consider the case of a
single flavour of quark with mass $m$; the flavour-conserving nature
of QCD makes the generalisation of the procedure to multiple flavours
straightforward.

To begin, we add a Dirac mass term to the original lagrangian, so
\begin{equation}
  \bar\psi \: i\slashed \gD \: \psi \rightarrow \bar\psi (i\slashed \gD - m)
  \psi
  \label{eq:}
\end{equation}
in \eqref{eq:lcqcd}. If we repeat the analysis of section
\ref{ssec:lcqcd}, we again arrive at a light-cone action given by
\eqref{eq:lcqcd-action} with the following additional terms in the
parentheses:
\begin{align}
  L^{\bar\psi\psi}_m &= \hphantom{-} \frac{im^2g^2}{4\sqrt 2}
  \int_\Sigma d^3\vec x (\bar\alpha^- \hat\partial^{-1} \alpha^++
  \bar\alpha^+ \hat\partial^{-1} \alpha^-),
  \label{eq:lcqcd-quarkmass} \\
  L^{\bar\psi^+-\psi^+} &= - \frac{mg^2}4 \int_\Sigma d^3\vec x \:
  \bar\alpha^+ [\hat\partial^{-1}, \bar\gA] \alpha^+,
  \label{eq:lcqcd-massint1} \\
  L^{\bar\psi^-+\psi^-} &= \hphantom{-} \frac{mg^2}4 \int_\Sigma
  d^3\vec x \: \bar\alpha^- [\hat\partial^{-1}, \gA] \alpha^-.
  \label{eq:lcqcd-massint2}
\end{align}
Note now that since $m\ne0$, chirality and helicity become mixed;
therefore the signs in the superscripts of the quark fields no longer
denote the corresponding out-going particles' helicities.  With the
term \eqref{eq:lcqcd-quarkmass}, the propagator for the light-cone
gauge action's quarks becomes
\begin{equation}
  \frac{i \sqrt 2 \: \hat p}{p^2 - m^2}.
\end{equation}
We note that the quark propagator conserves the chirality even in the
massive case, and that the only chirality violation comes from the
vertices in (\ref{eq:lcqcd-massint1}) and (\ref{eq:lcqcd-massint2}).

The polarisation spinors for massive states have at least three
non-zero components, and so couple to non-dynamical quark components
when applied in the LSZ reduction. To understand how we should deal
with this, let us study the gauge-fixing process with the source terms
instated. They contribute the following term to the action:
\begin{equation}
  S_{\text{src}} = \int d^4x \: \left\{ \frac{i\sqrt 2}g \tr J^\mu
    \gA_\mu + \bar\kappa \psi + \bar\psi\kappa \right\},
\end{equation}
where the fermionic sources are
\begin{equation}
  \kappa = (\kappa_1^+, \kappa_2^+, \kappa_2^-, \kappa_1^-)^{\rm T}
  \quad\text{and}\quad
  \bar\kappa = (\bar\kappa_2^+, \bar\kappa_1^+, \bar\kappa^-_1, \bar\kappa_2^-).
\end{equation}
Fixing $\hat\gA= 0$ and integrating out $\check\gA$ and the
non-dynamical fermions leaves
\begin{equation}
  \label{eq:lcqcd-src}
  \begin{split}
    S'_{\text{src}} = \int d^4x \: \Biggl\{ &\bar\kappa^+_1
    \hat\partial^{-1} \left( \bar\gD \alpha^+ - \frac{im}{\sqrt 2}
      \alpha^- \right) - \bar\kappa^-_1 \hat\partial^{-1} \left( \gD
      \alpha^- + \frac{im}{\sqrt 2} \alpha^+ \right)  \\
    &+ \hat\partial^{-1} \left( \gD \bar\alpha^- + \frac{im}{\sqrt 2}
      \bar\alpha^+ \right) \kappa^-_1 - \hat\partial^{-1} \left(
      \bar\gD \bar\alpha^+ - \frac{im}{\sqrt 2} \bar\alpha^- \right)
    \kappa^+_1 \\
    &+ \frac i{\sqrt 2} ( \bar\kappa^+_1 \hat\partial^{-1} \kappa^-_1
    +
    \bar\kappa^-_1 \hat\partial^{-1} \kappa_1^+ ) \\
    &+ \bar\kappa^+_2 \alpha^+ + \bar\kappa^-_2 \alpha^-
    + \bar\alpha^+ \kappa_2^+ + \bar\alpha^- \kappa_2^- \\
    &- \frac{i\sqrt 2}g \tr [\bar J \gA + J \bar\gA] + \text{$\hat J$
      terms} \Biggr\}.
  \end{split}
\end{equation}
The important thing to notice here, even before we have applied any
field transformation, is what has happened to the sources
$\kappa^\pm_1$ and $\bar\kappa^\pm_1$. These were previously coupled
to non-dynamical quark field components $\beta^\pm$ and
$\bar\beta^\pm$, but from \eqref{eq:lcqcd-src} we see that functional
derivatives of the partition function with respect to these sources
bring down insertions of a linear combination of the dynamical quark
degrees of freedom, and a completion vertex-like product of a gauge
and quark field. At the tree-level, the latter should be annihilated
for generic momenta as we take the on-shell limit in the LSZ
theorem. In contrast, we would initially expect the former to
contribute. However, as we shall see in section
\ref{ssec:mass-spinors}, we can choose our polarisation spinors so
that the non-dynamical components are once again decoupled.

Let us turn to the field transformation for the massive scenario.
First, suppose we hold on to much of the machinery of section
\ref{sec:transf}: specifically, we retain the definition of the
transformation in \eqref{eq:transf-def}, excluding the mass terms.
Since the definitions of the canonical momenta \eqref{eq:oldfields} do
not change, the calculation carries through as before, and we then
substitute the expressions we have obtained for the old fields into
\eqref{eq:lcqcd-quarkmass}--\eqref{eq:lcqcd-massint2}; the resulting
lagrangian will be the massless MHV lagrangian with additional terms
proportional to $m^2$ and $m$.\footnote{This method is also adopted in
  \cite{Boels:2007pj} and \cite{Boels:2008ef} in their discussion of
  massive scalars.}  This is the most straightforward way of extending
to the massive quark case from the massless one (we consider another
possible transformation just before section \ref{ssec:mass-spinors}).
Notice that the term of \eqref{eq:lcqcd-massint1} would appear to be
of the `wrong' form for a MHV lagrangian, but we must keep the
following in mind: it is of course no longer correct any more to talk
of the resulting terms as being of a maximally
\emph{helicity}-violating form since the quark fields in each term
only have definite chirality, not helicity, when masses are present.
Nevertheless, in each term the number of negative helicity gauge
fields is still constrained to be not more than two, and any given
term never contains any more than four quark fields, so for future
work on this topic it may be useful to classify vertices according to
their chirality content.

With these choices in place, the massive MHV lagrangian consists of:
the massless kinetic terms for the gluons and quarks (the mass terms
for the latter are discussed below); the massless QCD interaction
terms \eqref{eq:MHVL-massless-YM} and \eqref{eq:MHVL-massless-q}; and
now a new set of massive terms arising from the substitution of the
field transformation into
\eqref{eq:lcqcd-quarkmass}--\eqref{eq:lcqcd-massint2}. These new
massive pieces provide the MHV lagrangian with terms of the form
\begin{align}
  L_{\text{F}m}^{+-} &= \sum_{n=2}^\infty\int_{1\cdots n} \bigg\{
  V_{\text{F}m}^{-+}(1 \cdots n)\bar \xi_{\bar 1}^-{\cal B}_{\bar
    2}\cdots\gB_{\overline{n-1}}\xi^+_{\bar n}
  + V_{\text{F}m}^{+-}(1 \cdots n)\bar \xi_{\bar 1}^+\gB_{\bar
    2}\cdots{\cal B}_{\overline{n-1}}\xi^-_{\bar n} \bigg\},
  \label{eq:Lpm-mass-expansion}
  \\
  \begin{split}
    L^{+-+}_{\text{F}m} &=
    \sum_{n=3}^\infty\sum_{s=2}^{n-1}\int_{1\cdots
      n}V_{\text{F}m}^{s,+-+}(1\cdots n) \bar\xi_{\bar 1}^+ {\cal
      B}_{\bar 2}\cdots\bar\gB_{\bar s}\cdots{\cal
      B}_{\overline{n-1}}\xi^+_{\bar n}
    \\
    &\quad + \sum_{n=4}^\infty\sum_{s=2}^{n-2}\int_{1\cdots n} \bigg\{
    V_{\text{F}m}^{s,++-+}(1\cdots n) \bar\xi_{\bar 1}^+ \gB_{\bar
      2}\cdots{\cal B}_{\overline{s-1}}\xi_s^+\bar\xi^-_{\overline
      {s+1}} {\cal
      B}_{\overline{s+2}}\cdots\gB_{\overline{n-1}}\xi^+_{\bar n}
    \\
    &\hphantom{+ \sum_{n=4}^\infty\sum_{s=2}^{n-2}\int_{1\cdots n}
      \bigg\{} +V_{\text{F}m}^{s,+-++}(1\cdots n) \bar\xi_{\bar 1}^+
    {\cal B}_{\bar 2}\cdots\gB_{\overline{s-1}}\xi_{\bar
      s}^-\bar\xi^+_{\overline{s+1}} {\cal
      B}_{\overline{s+2}}\cdots\gB_{\overline{ n-1}}\xi^+_{\bar
      n}\bigg\},
  \end{split}
  \label{eq:Lpmp-mass-expansion}
  \\
  L^{-+-}_{\text{F}m} &= \sum_{n=3}^\infty\int_{1\cdots
    n}V_{\text{F}m}^{-+-}(1\cdots n) \bar\xi_{\bar 1}^- \gB_{\bar
    2}\gB_{\bar 3}\cdots\gB_{\overline{n-1}}\xi^-_{\bar n}.
  \label{eq:Lmpm-mass-expansion}
\end{align}
(The momentum-conserving $\delta$ function has been absorbed into the
integration measure as in \eqref{eq:MHVL-massless-YM} and
\eqref{eq:MHVL-massless-q}.)  It is straightforward though tedious to
calculate these vertices by substituting the series for $\gA$, $\bar
\gA$, $\alpha$, $\bar\alpha$ in terms of $\gB$, $\bar \gB$, $\xi$,
$\bar\xi$ into (\ref{eq:lcqcd-quarkmass})--(\ref{eq:lcqcd-massint2}):
\begin{align}
  V_{\text{F}m}^{-+} &= \hphantom{-}\frac{(-i)^{n-2}m^2g^2}{4\sqrt2}
  \frac{\hat 1\hat 2\cdots\widehat{n-1} \: (1\:n)} {\hat1\hat
    n(1\:2)(2\:3)\cdots(n-1,n)} \quad\text{for $n\ge 2$},
  \label{eq:massCSW-mp}\\
  V_{\text{F}m}^{+-} &= -\frac{(-i)^{n-2}m^2g^2}{4\sqrt2} \frac{\hat
    2\cdots\widehat{n-1} \: \hat n(1\:n)} {\hat1\hat
    n(1\:2)(2\:3)\cdots(n-1,n)} \quad\text{for $n\ge 2$},
  \label{eq:massCSW-pm}\\
  V_{\text{F}m}^{s,+-+} &= \hphantom{-}\frac{(-i)^{n-2}mg^2}4
  \frac{\hat 2\hat3\cdots\widehat{n-1} \: (1\:s)(s\:n)}{\hat1 \hat
    n(1\:2)(2\:3)\cdots(n-1,n)} \quad\text{for $n\ge 3$},
  \\
  V_{\text{F}m}^{-+-}&= -\frac{(-i)^{n-2}mg^2}4 \frac{\hat
    2\hat3\cdots\widehat{n-1} \: (1\:n)^2}{\hat1 \hat
    n(1\:2)(2\:3)\cdots(n-1,n)} \quad\text{for $n\ge 3$},
  \\
  \begin{split}
    V_{\text{F}m}^{s,++-+} &= \hphantom{-}\frac{(-i)^{n-2}mg^4}{8\sqrt
      2} \frac{\hat 2\hat3\cdots\widehat{n-1} \: (1\:s)}{ \hat1\hat
      s(1\:2)(2\:3)\cdots(n-1,n)}\bigg( \frac{(s+1,n)}{ \hat n }+\frac
    1
    {N_C} \frac {(s,s+1)}{ \hat s  }\bigg) \\
    &\qquad \text{for $n\ge 4$},\end{split}
  \\
  \begin{split}
    V_{\text{F}m}^{s,+-++} &= -\frac{(-i)^{n-2}mg^4}{8\sqrt 2}
    \frac{\hat 2\hat3\cdots\widehat{n-1} \: (s+1,n)}{\widehat{s+1}
      \hat n(1\:2)(2\:3)\cdots(n-1,n)} \bigg( \frac{(1\:s)}{
      \hat1}+\frac 1
    {N_C} \frac {(s,s+1)}{ \widehat{ s+1} }\bigg) \\
    &\qquad \text{for $n\ge 4$}.\end{split}
\end{align}
(Note above that we have dropped the argument list $(1\cdots n)$ from
the left-hand sides.)  Similarly to the amplitudes, the vertices have
the relations:
\begin{align*}
  V^{+-}_{\text{F}m}(1\cdots n) &=
  (-1)^{n+1}V^{-+}_{\text{F}m}(n\cdots1), \\
  V^{s,+-+}_{\text{F}m}(1\cdots n) &=
  (-1)^{n+1}V^{n-s,+-+}_{\text{F}m}(n\cdots1), \\
  V^{s,++-+}_{\text{F}m}(1\cdots n) &=
  (-1)^{n}\hphantom{{}^{+1}}V^{n-s,+-++}_{\text{F}m}(n\cdots1), \\
  V^{-+-}_{\text{F}m}(1\cdots n) &=
  (-1)^{n+1}V^{-+-}_{\text{F}m}(n\cdots1).
\end{align*}
Equations \eqref{eq:massCSW-mp} and \eqref{eq:massCSW-pm} are
proportional to the CSW vertices for massive scalars as derived in
\cite{Boels:2007pj}, which may imply a supersymmetric relation between
these two kinds of vertices.

Since to leading order $\alpha$ is just $\xi$, the leading order terms
of the expansion proportional to $m^2$ in the MHV lagrangian (\ie\ the
$n=2$ terms in \eqref{eq:Lpm-mass-expansion}) are obtained by
replacing $\alpha$ with $\xi$ in \eqref{eq:lcqcd-quarkmass}.  Also,
the series expansions are also the expansion with respect to $g$, a
factor of which is hidden in each $\gB$. If we consider the
canonically normalised form of the action, including the $4/g^2$
factor from \eqref{eq:lcqcd-action}, we see that the terms in
\eqref{eq:Lpm-mass-expansion} form a series in $g^{n-2}$ for $n \ge
2$.  Thus we merge the $n=2$ terms in (\ref{eq:Lpm-mass-expansion})
into the kinetic part on which the perturbative expansion is based,
and the higher order expansion terms into the interaction part.  In
this way, the propagator for the new fermion fields is just the same
as that of the old ones:
\begin{equation}
  \langle \xi^\pm \bar\xi^{\mp}\rangle=
  \frac{i\sqrt2\: \hat p}{p^2-m^2}\,, 
\end{equation} 
which is just the corresponding component of the ordinary propagator
$i(\slashed p+m)/(p^2-m^2)$.  Similarly, we see that the terms that
\eqref{eq:lcqcd-massint1} and \eqref{eq:lcqcd-massint2} supply to the
MHV lagrangian, which are also proportional to $m$, generate a
perturbation series in $g$ starting at ${\cal O}(g)$. In section
\ref{ssect:masses-qpq}, as a sample calculation, we will check the
three-particle amplitudes obtained from the MHV lagrangian against
those from ordinary massive light-cone gauge QCD.

Parenthetically, we note that one might choose another way of
generalising to massive quarks, specifically including the massive
terms in the transformation equation according to
\begin{multline}
  L^{-+}[\gA, \bar\gA] + L^{-++}[\gA, \bar\gA] +
  L^{\bar\psi\psi}[\alpha^\pm,\bar\alpha^\pm] +
  L^{\bar\psi+\psi}[\gA,\alpha^\pm,\bar\alpha^\pm] +
  L^{\bar\psi\psi}_m[\alpha^\pm,\bar \alpha^\pm] \\
  + L^{\bar\psi^+-\psi^+}[\alpha^+,\bar\gA] = L^{-+}[\gB, \bar\gB] +
  L^{\bar\psi\psi}[\xi^\pm,\bar\xi^\pm]+
  L^{\bar\psi\psi}_m[\xi^\pm,\bar \xi^\pm] .
  \label{eq:transf-def-masses}
\end{multline}
The difficulty with such a transformation is that there is no reason
to expect a simple closed form for the coefficients of the expansion
in the new fields any more (something similar was seen in ref.\
\cite{Ettle:2007qc}, when considering the pure Yang--Mills MHV
lagrangian outside four dimensions). On the other hand, one may be
able to derive closed forms for the coefficients of the expansion in
the new fields if one also simultaneously expands in an infinite
series of higher powers in $m$. This alternative approach is clearly
more complicated than the one we pursue here, in which the expansion
coefficients have a simple elegant closed form independent of the
quark mass, and the lagrangian only has terms with powers of the quark
mass not exceeding two.
 
\subsection{Massive polarisation spinors}
\label{ssec:mass-spinors}

The Dirac equations in momentum space are
\begin{equation}
  (\slashed p -m)u^\sigma(\vec p)=0
  \quad\text{and}\quad
  (-\slashed p -m)v^\sigma(\vec p)=0\,,
\end{equation}
where $u$ and $v$ are the positive- and negative-energy solutions,
respectively.  Now, for the purposes of the LSZ reduction, we can
choose our polarisation spinors to satisfy these equations, but
instead with a RHS that vanishes only on shell. Furthermore, we can
choose this RHS to have components that result in $\beta^\pm$ and
$\bar\beta^\pm$ being decoupled from the reduction as in equations
\eqref{eq:decouple-massless} and \eqref{eq:invprop} in massless cases,
\ie\ we search for off-shell solutions $u^\sigma$ and $v^\sigma$ that
satisfy
\begin{equation}
  (\slashed p -m)u^\sigma(p) =
  \begin{pmatrix} 0 \\ * \\ * \\ 0 \end{pmatrix} (p^2-m^2),
  \quad
  (\slashed p+m)v^\sigma(p) =
  \begin{pmatrix} 0 \\ * \\ * \\ 0 \end{pmatrix} (p^2-m^2).
  \label{eq:offshellreq}
\end{equation}
where the asterisks denote unspecified quantities.

Let us concentrate on the positive-energy spinor $u$ and postulate
solutions of the form
\begin{equation}
  u^{\pm}= \begin{pmatrix} a^{\pm}\eta_1^{\pm} \\ b^{\pm}\eta_2^{\pm} \end{pmatrix}
  \label{eq:ansatz}
\end{equation}
where
\begin{eqnarray}
  \eta_1^{+}=\eta^{+}
  &\text{and}&\,
  \eta_2^{+} = \eta^{+} + \rho^{+}
  \begin{pmatrix} p^2-m^2 \\0 \end{pmatrix}\,,
  \label{eq:helicity-bases-mass1}  \\
  \eta_2^{-}=\eta^{-}\,
  &\text{and}&\,
  \eta_1^{-} = \eta^{-} + \rho^{-}
  \begin{pmatrix} 0\\  p^2-m^2 \end{pmatrix}\,.
  \label{eq:helicity-bases-mass2} \end{eqnarray}
Here, the $\eta^\pm$ are the eigenspinors of the helicity operator and
solve the equation
\begin{equation}
  \frac{\vec p \cdot \pmb\sigma}{|\vec p|}\eta^\lambda = \lambda \eta^\lambda.
\end{equation}
This has two solutions for $\lambda=\pm$,
\begin{equation}
  \eta^{+}=-\sqrt2\Big(p,p^--\hat p\Big)^{\rm T}\,,
  \quad\text{and}\quad
  \eta^{-}=-\sqrt2\Big(p^--\hat p,-\bar p\Big)^{\rm T}\,,
\end{equation}
in which we have defined
\begin{equation}
  p^{\pm} := \frac1{\sqrt2}(p^t \pm |\vec p|)
\end{equation}
in order to tidy up notations in the following equations.  In the
massless limit, $\eta^{\pm}$ recovers $\varphi$, $\bar\omega$ in
(\ref{eq:poln-aqm}) and (\ref{eq:poln-aqp}) up to a normalization.
 
By solving \eqref{eq:offshellreq}, and using the normalization
$u^\dagger u=2p^t$, the $u^{\pm}$ are
\begin{align}\label{eq:u-massive-p}
  u^{+}(p)&=\nu
  \begin{pmatrix}
    -\sqrt2 m p\cr -\sqrt2m(p^{-}-\hat p)\cr -p
    \big[m^2-2p^{+}(p^{-}-\hat p)\big]/\hat p \cr -2p^{+}(p^{-}-\hat
    p)
  \end{pmatrix}\,, \\ \quad\text{and}\quad u^{-}(p)&=\nu
  \begin{pmatrix}
    -2p^{+}(p^{-}-\hat p) \cr \bar p \big[m^2-2p^{+}(p^{-}-\hat
    p)\big]/\hat p \cr -\sqrt2m(p^{-}-\hat p) \cr \sqrt 2 m \bar
    p\end{pmatrix}\,,
  \\
  \text{where}\quad \nu&=\frac1{2\sqrt{|\vec p|p^{+}(\hat
      p-p^{-})}}\,.
  \label{eq:u-massive-m}\end{align}
The normalisation $\nu$ is just an overall coefficient and does not
affect the off-shell condition of \eqref{eq:offshellreq}. We have
therefore set $p^2 = m^2$ within it to simplify the result.  It is
also easy to verify that in the massless limit the solution reduces to
the previous massless $u^\pm$ using the limiting behavior:
\begin{eqnarray}
  p^+\to \sqrt 2 p^t\,,\quad p^-\to \frac{m^2}{2\sqrt2 p^t}\,,
  \quad \hat p-p^{+}\to -\check p\,,\quad \hat p-p^-\to \hat p
  \,,\quad \nu\to \frac 1{2^{5/4}p^t\sqrt{\hat p}}\,.
\end{eqnarray}

For the antiquark, we replace $m \rightarrow -m$ and flip the helicity
in above equations, \ie\ $v^\pm=u^\mp|_{m\to -m}$.  If we define:
\begin{eqnarray}
  \bar\phi^{\dot\alpha}=\nu
  \begin{pmatrix}
    -\sqrt2 m p\cr -\sqrt2m(p^{-}-\hat p)
  \end{pmatrix}
  \quad \text{and}\quad
  \chi_{\alpha}=\nu
  \begin{pmatrix}
    - p \big[m^2-2p^{+}(p^{-}-\hat p)\big]/\hat p \cr
    -2p^{+}(p^{-}-\hat p)
  \end{pmatrix}\,,
\end{eqnarray}
$u^{\pm}$ and $v^{\pm}$ can be represented in a more compact form,
where we have used $\epsilon^{\dot\alpha\dot\beta}$ and
$\epsilon_{\alpha\beta}$ to raise and lower spinor indices:
\begin{eqnarray}
  u^+=\begin{pmatrix} \bar\phi^{\dot\alpha}\cr \chi_{\alpha}\end{pmatrix},
  \quad
  u^-=\begin{pmatrix} \bar\chi^{\dot\alpha}\cr -\phi_{\alpha}\end{pmatrix},
  \quad
  v^{+}=\begin{pmatrix} \bar\chi^{\dot\alpha}\cr  \phi_{\alpha}\end{pmatrix},
  \quad
  v^{-}=\begin{pmatrix}- \bar\phi^{\dot\alpha}\cr  \chi_{\alpha}\end{pmatrix}.
\end{eqnarray}
Explicit calculation shows that these solutions are orthogonal for
off-shell $p$, \ie\ $\bar u^+(p)\gamma^0 u^-(p) = \bar u^{+}(p) u^-(p)
= \bar u(p) v(p) = 0$.

Consequently, \eqref{eq:offshellreq} becomes
\begin{align}
  (\slashed p -m)u^+(p) &= \frac{\nu}{\hat p}
  \begin{pmatrix}
    0 \\ -\sqrt2p^+(p^{-}-\hat p) \\ -m p \\ 0
  \end{pmatrix}
  (p^2-m^2) \,,
  \label{eq:off-shell-Dirac1}  \\
  (\slashed p -m)u^-(p) &= \frac{\nu}{\hat p}
  \begin{pmatrix}
    0 \\m \bar p \\ -\sqrt2p^+(p^{-}-\hat p) \\ 0
  \end{pmatrix}
  (p^2-m^2) \,,
  \label{eq:off-shell-Dirac2}  \\
  (\slashed p +m)v^+(p) &= \frac{\nu}{\hat p}
  \begin{pmatrix}
    0 \\ -m \bar p\\
    -\sqrt2p^+(p^{-}-\hat p) \\ 0
  \end{pmatrix}
  (p^2-m^2) \,,
  \label{eq:off-shell-Dirac3}  \\
  (\slashed p +m)v^-(p) &= \frac{\nu}{\hat p}
  \begin{pmatrix}
    0 \\
    -\sqrt2p^+(p^{-}-\hat p) \\ m p \\ 0
  \end{pmatrix}
  (p^2-m^2) \,.\label{eq:off-shell-Dirac4}
\end{align}
Let us compare these solutions with the off-shell definition of the
spinor from refs. \cite{Kleiss:1985yh, Ballestrero:1994jn,
  Rodrigo:2005eu, Schwinn:2006ca}, which take
\begin{eqnarray}
  u(p,\pm)=\frac{\slashed p+m}{\langle
    p^\flat\mp|q{\pm}\rangle}|q\pm\rangle\,,
  \label{eq:Spinor-BS}
\end{eqnarray}
where $|q\pm\rangle$ is a Weyl spinor for an arbitrary null vector
$q$, and $p^\flat$ is the null projection of $p$ obtained by
subtracting a vector proportional to $q$.  If we choose $|q\pm\rangle$
to be the eigenspinor of helicity $h=\vec p\cdot \pmb \sigma/|\vec
p|$, this spinor has also definite helicity off shell, since $h$
commutes with $\slashed p+m$.  Due to the additional term proportional
to $p^2-m^2$ in \eqref{eq:helicity-bases-mass1} and
\eqref{eq:helicity-bases-mass2}, our choices of spinors are not
helicity eigenspinors off shell, so our off-shell continuation of the
spinors is different from \eqref{eq:Spinor-BS}. Hence our spinors
cannot be written in that form. This can also be seen from
\eqref{eq:off-shell-Dirac1}--\eqref{eq:off-shell-Dirac4}: if we
multiply \eqref{eq:Spinor-BS} by $\slashed p-m$, the right hand side
is $p^2-m^2$ multiplied with a Weyl spinor $|q\pm\rangle$, but in
\eqref{eq:off-shell-Dirac1}--\eqref{eq:off-shell-Dirac4}, it is
obvious that the right hand side is not a Weyl spinor.

However, we can establish a relation between our spinors and
\eqref{eq:Spinor-BS}: in
\eqref{eq:ansatz}--\eqref{eq:helicity-bases-mass2} the spinors
\emph{without} the additional off-shell term have definite helicity,
and these can be cast in the form of \eqref{eq:Spinor-BS}. Thus, our
choice simply adds an extra term to \eqref{eq:Spinor-BS}:
\begin{eqnarray}
  u(p,+)=\frac{\slashed p+m}{\langle p^\flat+|q{-}\rangle}|q-\rangle+
  \nu \frac p
  {\hat p}(p^2-m^2)|\mu,+\rangle,
  \label{eq:u-p-compare}  \\
  u(p,-)=\frac{\slashed p+m}{\langle p^\flat-|q{+}\rangle}|q+\rangle+ \nu
  \frac {\bar p}
  {\hat p}(p^2-m^2)|\mu,-\rangle,
  \label{eq:u-m-compare}
\end{eqnarray}
where $\mu=(1,0,0,1)/2$ in Minkowski co-ordinates (with spinors
$|\mu+\rangle=(0,0,1,0)^{\rm T}$ and $|\mu-\rangle=(0,-1,0,0)^{\rm
  T}$) and $q=(|\vec p|,-\vec p)$ are null vectors.  The specific
choice of $q$ here is such that it makes the $\pm$ embellishment of
the spinor $u$ coincide with the physical helicity of the fermion; for
arbitrary $q$, this embellishment loses this physical significance.

Let us now use the above equations to make the application of our
solutions to the LSZ formalism concrete. For an out-going quark,
\begin{align}
  \begin{split}
    \bar u^{+}(p)\,i(-\slashed p &+m) \langle \cdots
    \psi(p)\cdots\rangle
    \\
    &= \nu \Big(-\sqrt 2 m\bar p {\cal V}(\cdots
    \bar\alpha^{-}(-p)\cdots) - 2p^{+}(p^--\hat p){\cal V}(\cdots
    \bar\alpha^{+}(-p)\cdots) \Big),
  \end{split}
  \label{eq:LSZ-baru-p}
  \\
  \begin{split}
    \bar u^{-}(p)\,i(-\slashed p & +m) \langle \cdots
    \psi(p)\cdots\rangle
    \\
    & = \nu \Big(\sqrt 2 m p {\cal V}(\cdots \bar\alpha^{+}(-p)\cdots)
    - 2p^{+}(p^--\hat p){\cal V}(\cdots \bar\alpha^{-}(-p)\cdots)
    \Big),
  \end{split} \label{eq:LSZ-baru-m}
\end{align}
respectively for positive- and negative-helicity, where ${\cal V}$ is
the amputated correlation function computed using the light-cone or
MHV action. Similarly, for an out-going antiquark,
\begin{align}
  \begin{split}
    \langle \cdots \bar\psi(p)\cdots\rangle & i(-\slashed
    p-m)v^{+}(p)  \\
    & =\nu\Big(-\sqrt 2 m \bar p{\cal V}(\cdots \alpha^-(-p)\cdots)-2
    p^+(p^--\hat p) {\cal V}(\cdots \alpha^+(-p)\cdots) \Big),
  \end{split}
  \label{eq:LSZ-v-p}
  \\
  \begin{split}
    \langle \cdots \bar\psi(p)\cdots\rangle & i(-\slashed p-m)v^{-}(p)
    \\ & =\nu\Big(\sqrt 2 m p{\cal V}(\cdots \alpha^+(-p)\cdots)-2
    p^+(p^--\hat p) {\cal V}(\cdots \alpha^-(-p)\cdots) \Big).
  \end{split}
  \label{eq:LSZ-v-m}
\end{align}
Note that for definite helicity external quarks, both quark
chiralities contribute to the amplitude.

Considering \eqref{eq:u-massive-p}--\eqref{eq:u-massive-m}, we see
that the external factors in \eqref{eq:LSZ-baru-p}--\eqref{eq:LSZ-v-m}
are just the components of the polarisation spinors that correspond to
$\alpha$ and $\bar\alpha$. Since the additional terms in
(\ref{eq:helicity-bases-mass1}) and (\ref{eq:helicity-bases-mass2}) do
not affect these components, the external factors are the same as the
components of the spinor solutions without the additional off-shell
terms.  So we conclude that one does not need to consider
contributions from the non-dynamical quark components to the
amplitudes in light-cone QCD or MHV lagrangian-based calculations. As
a check of this decoupling of $\beta^\pm$ and $\bar\beta^\pm$, in
section \ref{ssect:masses-qmq} we will compare the three point
off-shell amplitude $A(1_{\rm q}^\pm2^-3_{\rm \bar q}^\pm)$ calculated
using light-cone gauge QCD with the one calculated directly from
Feynman gauge QCD. It is worth pointing out that this discussion does
not depend on the specific choice of the helicity basis $\eta^\pm$ in
\eqref{eq:helicity-bases-mass1} and \eqref{eq:helicity-bases-mass2};
one can also choose a different $q$ for \eqref{eq:u-p-compare} and
\eqref{eq:u-m-compare} and a different coefficient before the factor
$p^2-m^2$, repeat the foregoing, and arrive at the same conclusion.

\subsection{Helicity flipping property of the amplitude}
\label{ssect:Helicity-flipping}

We can now look at what happens to the amplitude when we flip the
external helicities. For amplitudes of massless QCD, we know that
under such an operation, the corresponding amplitude exchanges the
holomorphic and anti-holomorphic variables. We will show that in the
amplitude with massive quarks, this property should be modified a
little: we need to multiply the result by $(-1)^f$ where $f=\tfrac12
(n_+ - n_-)$. Here, $n_\pm$ is the total number of positive (negative)
helicity quarks and antiquarks.

For a quark-antiquark pair, there could be four kinds of terms in the
amplitude: terms composed of the amputated correlation functions with
$\bar\alpha^+ \alpha^+$, $\bar\alpha^+ \alpha^-$,
$\bar\alpha^-\alpha^+$, $\bar\alpha^- \alpha^-$ and corresponding
coefficients from LSZ formulae
\eqref{eq:LSZ-baru-p}--\eqref{eq:LSZ-v-m}.  From the first two
equations, we can see when the helicity of the quark is flipped from
$+$ to $-$, the factors corresponding to $\bar\alpha^-$ changes to the
one corresponding to $\bar\alpha^+$ by exchanging the the holomorphic
variable and anti-holomorphic variables and changing the overall sign,
{\it i.e.} $-\sqrt2 m \bar p\to \sqrt2 m p$, whereas the term with
$\bar \alpha^+\to \bar\alpha^-$ exchanges the holomorphic and
anti-holomorphic variables without changing the sign. And when the
helicity changes from $-$ to $+$, the factors behave in just the
opposite fashion. For the antiquark, there is a similar property. This
has been summarised in table \ref{tbl:flip-1}.

\TABLE{
  \begin{tabular}{ccc}
    \toprule
    &     $\bar\alpha^+\to\bar \alpha^-$& $\bar \alpha^-\to\bar \alpha^+$\cr
    \midrule
    $q^+\to q^-$             &   $+$   &   $-$   \cr
    $q^-\to q^+$             &   $-$   &   $+$   \cr
    \bottomrule
  \end{tabular}\quad
  \begin{tabular}{ccc}
    \toprule
    &     $\alpha^+\to \alpha^-$& $\alpha^-\to \alpha^+$\cr
    \midrule
    $\bar q^+\to \bar q^-$   &   $+$   &   $-$   \cr
    $\bar q^-\to \bar q^+$   &   $-$   &   $+$   \cr
    \bottomrule
  \end{tabular}
  \label{tbl:flip-1}
  \caption{The sign changes that follow from the LSZ formulae of
    \eqref{eq:LSZ-baru-p}--\eqref{eq:LSZ-v-m}.}
}

As a result, the flipped amplitude can be obtained from the original
amplitude by the following operations: changing chirality of
corresponding $\bar \alpha \alpha$ pair and helicity of $A$ of the
amputated correlation function, exchanging the holomorphic and
anti-holomorphic variables of corresponding external quark factor and
at the same time making the sign changes in corresponding terms as
table \ref{tbl:flip-2} for each quark-antiquark pair.

\TABLE{
  \begin{tabular}{ccc}
    \toprule
    &$\bar\alpha^+\alpha^+\leftrightarrow\bar\alpha^-\alpha^-$
    &$\bar\alpha^+\alpha^-\leftrightarrow\bar\alpha^-\alpha^+$ \\
    \midrule
    $q^+\bar q^+\leftrightarrow q^-\bar q^-$   &   $+$   &    $-$
    \cr
    $q^-\bar q^+\leftrightarrow q^+\bar q^-$   &   $-$   &    $+$ \\
    \bottomrule
  \end{tabular}
  \label{tbl:flip-2}
  \caption{The corresponding sign changes for quark-antiquark pairs
    that follow from table \ref{tbl:flip-1}.}
}

Moreover, the two chirality violating terms in lagrangian
(\ref{eq:lcqcd-massint1}) and (\ref{eq:lcqcd-massint2}) swap when we
exchange the holomorphic and anti-holomorphic variables, flip the
chirality and change the overall sign, while the other chirality
conserving vertices in lagrangian from (\ref{eq:lcqcd-qbarq}) to
(\ref{eq:lcqcd-qbarqqbarq}) do not change sign under this operation.
This means that for an external like-chirality $\alpha^\pm\bar
\alpha^\pm$ pair connected by an internal quark line, the operation
corresponding to the first column in the above table contributes a
minus sign to the amputated correlation function after we exchange the
holomorphic and anti-holomorphic variables, because there should be an
odd number of chirality violating vertices on the fermion line
connecting these two fermions. Multiplying the first column with this
sign change, we arrive at the sign changes we need to apply for each
quark-antiquark pair, after exchanging all holomorphic and
anti-holomorphic variables.  So we can see that for a like-helicity
quark-antiquark pair there will be a minus sign.  It is clear that
there is an even (odd) number of like-helicity quark antiquark pairs
when $f$ is even (odd), and thus our original assertion is proved.

\subsection{The three point amplitude $A(1_{\rm q}^{\pm} 2^- 3_{\bar
    {\rm q}}^{\pm})$}
\label{ssect:masses-qmq}

In this section we will compare the amplitude $A(1_{\rm q}^{\pm} 2^-
3_{\bar {\rm q}}^{\pm})$ calculated using the light cone lagrangian
combined with LSZ formulae \eqref{eq:LSZ-baru-p}--\eqref{eq:LSZ-v-m},
with the one computed using Feynman gauge QCD using the same $\bar
u(p)$ and $v(p)$ from \eqref{eq:u-massive-p}--\eqref{eq:u-massive-m}.
As we noted earlier, this is in order to check the conclusion in
section \ref{ssec:mass-spinors} that one does not need to consider the
contribution of the non-dynamical quark variables to the amplitudes
when using light-cone QCD and MHV lagrangian methods.  Notice here
that the notation $A(1_{\rm q}^{\pm}2^-3_{\bar {\rm q}}^{\pm})$
denotes \emph{four} amplitudes (not two), \ie\ the quark helicity
superscripts are independent.

In Feynman gauge QCD the only interaction term which contributes to
the amplitude is $\frac g{\sqrt{2}}\bar\psi \slashed{A} \psi$ and the
amplitude can be read out directly:
\begin{eqnarray}
  A(1_{\rm q}^{\pm}2^-3_{\bar {\rm q}}^{\pm})
  = -i \frac g {\sqrt2}\bar u^\pm_1 \slashed A_2^- v^\pm_3,
\end{eqnarray}
the minus sign coming from the fermion statistics.  Let us look at
$A(1_{\rm q}^{+}2^-3_{\bar {\rm q}}^{-})$ first: by using the explicit
expression of $\slashed A^-$ with $\mu=(1,0,0,1)/\sqrt 2$ as the
reference momentum in the gluon's polarisation vector,
\begin{eqnarray}
  \slashed A^-=
  \sqrt 2\begin{pmatrix}
    & & 0 & 0\\
    &&-1& -k/\hat k\\
    -k/\hat k&0&&\\
    1&0&&
  \end{pmatrix}\,,
\end{eqnarray}
(where the missing entries are all $0$, omitted for clarity) and
equation (\ref{eq:u-massive-p}), we eventually obtain:
\begin{equation}
  \begin{split}
    A(1_{\rm q}^{+}2^-3_{\bar {\rm q}}^{-})=-ig\nu_1\nu_3
    \bigg[2m^2\Big(\frac {\bar 1 \tilde 2\tilde3}{\hat
      2}-\frac{\tilde3}{\hat 3} (1^--\hat 1)(\hat 3+1^{+})\Big)
    \\
    + 4 \frac{1^{+}3^{+}(2\:3)}{\hat 2 \hat 3} (1^- -\hat 1)(3^- -\hat
    3)\bigg]\,.
  \end{split}\label{eq:QCD-massive-pmm}
\end{equation}
Notice that we have only used momentum conservation here, but not the
on-shell condition in the calculation, since the three external
particles cannot be all on shell at the same time.  One can easily
check that when $m\to 0$, this amplitude recovers the massless
off-shell amplitude.

Let us repeat this exercise using massive light-cone QCD, the terms of
which that contribute to the amplitudes are (\ref{eq:lcqcd-qbarmq})
and (\ref{eq:lcqcd-massint1}).  Reformulating them in momentum space
\and now converting to the canonical normalisation (\ie\ such that the
action is normalised as $S=\int d^4x\: \sum \tilde L$, \cf\
\eqref{eq:lcqcd-action}) for convenience:
\begin{align}
  \label{eq:massive-mhv-1}
  \tilde L^{\bar\psi-\psi} &= g\bigg[ \frac{(1\:2)}{\hat 1\hat 2}\bar
  \alpha^-_{\bar1} \bar A_{\bar2} \alpha^+_{\bar3} -\frac{(2\:3)}{\hat
    2 \hat3}\bar \alpha^+_{\bar1} \bar A_{\bar2}
  \alpha^-_{\bar3}\bigg],
  \\
  \label{eq:massive-mhv-2}
  \tilde L^{\bar\psi^+-\psi^+} &= \frac{gm}{\sqrt2} \frac{\hat
    2}{\hat1\hat3}\bar \alpha^+_{\bar1}\bar A_{\bar2}
  \alpha^+_{\bar3}\,.
\end{align}
After extracting the vertices from the above two equations and using
equations (\ref{eq:LSZ-baru-p}) and (\ref{eq:LSZ-v-m}) we can write
down:
\begin{multline}
  A(1_{\rm q}^{+}2^-3_{\bar {\rm q}}^{-})= -i\nu_1\nu_3\bar E_2\bigg[
  \sqrt2m\bar 1\times\left(g\frac{(1\:2)}{\hat 1\hat
      2}\right)\times(-\sqrt2m\tilde3)
  \\
  + 4 \times 1^{+}(\hat1 -1^{-})\times\left(-g\frac{(2\:3)}{\hat 2
      \hat3}\right) \times (\hat 3-3^{-})3^{+}
  \\
  +2 \times 1^{+}(\hat1-1^{-})\times\left(\frac{gm}{\sqrt2} \frac{\hat
      2}{\hat1\hat3} \right)\times(- \sqrt2 m \tilde 3)\bigg]\,,
\end{multline}
where $\bar E_2=-1$ is the polarization of $\bar A_2$.  The second
term above is just the second term of \eqref{eq:QCD-massive-pmm}, and
using the fact that $p\bar p=-(\hat p-p^+)(\hat p-p^-)$, we see that
the first and third terms above sum to the first term in
\eqref{eq:QCD-massive-pmm}.  This validates (at least in this
particular case) the use of the Feynman rules derived from the
light-cone QCD action, when combined with the LSZ reduction as in
\eqref{eq:LSZ-baru-p}--\eqref{eq:LSZ-v-m}. Furthermore, we note that
this also validates the use of the Feynman rules obtained from MHV
lagrangian as well, since \eqref{eq:massive-mhv-1} and
\eqref{eq:massive-mhv-2} both correspond to terms found therein by
leading-order substitutions into \eqref{eq:lcqcd-qbarmq} and
\eqref{eq:lcqcd-massint1}, respectively.

Similarly, we can also calculate the other three amplitudes:
\begin{eqnarray}
  &&\begin{aligned}
    A(1_{\rm q}^{-}2^-3_{\bar {\rm q}}^{+})=-ig\nu_1\nu_3
    \bigg[&2m^2\Big(\frac {\tilde 1 \tilde 2\bar3}{\hat 2}-\frac{\tilde1}{\hat 1}
    (3^--\hat 3)(\hat 1+3^{+})\Big)
    \\
    &- 4 \frac{1^{+}3^{+}(1\:2)}{\hat 1 \hat 2}
    (\hat 1-1^{-})(\hat 3-3^{-})\bigg]\,,
  \end{aligned}\label{eq:QCD-massive-mmp}
  \\
  &&\begin{aligned}
    A(1_{\rm q}^{+}2^-3_{\bar {\rm q}}^{+})=-ig\nu_1\nu_3
    \bigg[2\sqrt2 m\Big(
    &-\frac {\bar 1 \tilde 2 3^+}{\hat 2}(3^{-}-\hat 3)
    -\frac {1^+ \tilde 2 \bar3}{\hat 2}(1^{-}-\hat 1)
    \\&+ (1^--\hat1)(3^--\hat 3)( 1^++3^{+})\Big)
    \bigg]\,,
  \end{aligned}\label{eq:QCD-massive-pmp}
  \\
  &&\begin{aligned}
    A(1_{\rm q}^{-}2^-3_{\bar {\rm q}}^{-})=-ig\nu_1\nu_3
    \bigg[& 
    2\sqrt2 m\Big(\frac {1^+ \tilde 3\,(1\:2)}{\hat1\hat 2}(1^{-}-\hat 1)
    -\frac {\tilde1 3^+(2\:3)}{\hat 2\hat3}(3^{-}-\hat 3)\Big)
    \\&-\sqrt2 m^3 \frac{\tilde1\hat2\tilde3}{\hat1\hat3}
    \bigg]\,.
  \end{aligned}\label{eq:QCD-massive-mmm}
\end{eqnarray}
We can see that $A(1_{\rm q}^{+}2^-3_{\bar {\rm q}}^{+})$ and
$A(1_{\rm q}^{-}2^-3_{\bar {\rm q}}^{-}) $ are symmetric under
$1\leftrightarrow3$ and vanish in the massless limit.

\subsection{The three point amplitude: $A(1_{\rm q}^{\pm} 2^+ 3_{\bar
    {\rm q}}^{\pm})$}
\label{ssect:masses-qpq}

In this section, we will compare the amplitudes $A(1_{\rm
  q}^{\pm}2^+3_{\bar {\rm q}}^{\pm})$ calculated using light-cone
gauge massive QCD with those from the massive MHV lagrangian in order
to check that our massive MHV lagrangian scenario is valid and
equivalent to ordinary perturbation method.
 
In light-cone QCD, we would expect contributions to this amplitude
from \eqref{eq:lcqcd-qbarpq} and \eqref{eq:lcqcd-massint2}. On the
other hand, in the MHV lagrangian \eqref{eq:lcqcd-qbarpq} has been
eliminated by the transformation, but the amplitude will receive
contributions from the leading-order substitutions for the new fields
into \eqref{eq:lcqcd-massint2}, and the next-to-leading-order terms of
the transformation into \eqref{eq:lcqcd-quarkmass}.  We should still
keep in mind that there are MHV completion vertex contributions which
should be taken into account since we are dealing with the off-shell
amplitude.  Therefore the sum of the contributions from both
\eqref{eq:lcqcd-quarkmass} and completion vertices on the MHV
lagrangian side should be equal to the one from
\eqref{eq:lcqcd-qbarpq} in light-cone QCD.

First, consider the ${\cal O}(B)$ terms in the expansion of
\eqref{eq:lcqcd-quarkmass} in terms of the new fields (again, changing
to the canonical normalisation):
\begin{eqnarray}\label{eq:lcqcd-quark-m1-p}
  \tilde L_m^{\bar\psi\psi}=
  \frac {m^2 g}2\left (\frac{\hat2}{(1\:2)\hat3} \bar \xi^-_{\bar1}
    B_{\bar2}\xi^+_{\bar3}
    -\frac{\hat 2 }{(1\:2)\hat1}\bar \xi^+_{\bar1} B_{\bar2}\xi^-_{\bar3}\right)\,.
\end{eqnarray}
As an example, let us look at $A(1_{\rm q}^{+}2^+3^{-}_{\bar{\rm
    q}})$.  From the LSZ reduction formula (\ref{eq:LSZ-baru-p}) and
(\ref{eq:LSZ-v-m}), we can write down the contribution to the
amplitude from the vertex in the above term:
\begin{multline}
  i\frac{m^2g}2 \nu_1\nu_3 \bigg[ (\sqrt2 m\bar 1)\times\left
    (\frac{\hat2}{(1\:2)\hat3}\right) \times (-\sqrt2 m\tilde3)
  \\
  +4\times1^{+}(1^{-}-\hat 1)\times \left(-\frac{\hat 2
    }{(1\:2)\hat1}\right)\times(3^{-}-\hat 3)3^{+})\bigg]\,.
  \label{eq:MHV-qpq-m}
\end{multline}
Next we calculate the MHV completion vertices' contribution to the
correlation functions $\langle\alpha_{1}^+\bar A_{2}\bar \alpha_{
  3}^{-}\rangle$ and $\langle\alpha_{1}^-\bar A_{2}\bar \alpha_{
  3}^+\rangle$ and use the LSZ formula.  We expand the fields to
next-to-leading order, taking the $B\xi$ term for $\alpha_1$, the
$\xi\bar\xi$ term for $\bar B_2$, and the $\bar\xi B$ term for
$\bar\alpha_3$ to obtain:
\begin{align}
  \begin{split} \Big\langle\alpha_{1}^+{ }_{i}\:\bar A^{a_2}_{2}\bar
    \alpha_{ 3}^{-\bar \jmath}\Big\rangle &= -\frac g{\sqrt 2}\frac
    {\hat 2+\hat 3}{(2\:3)}\Big\langle
    \big(B_{\bar2}\xi_{\bar3}^+\big)_i\:\bar B_{2}^{a_2}\bar
    \xi_{3}^{-\bar\jmath}\Big\rangle -\frac g{\sqrt2}\frac {\hat
      1}{(1\:2)}\Big\langle \xi_{ 1 }^+{ }_i\:\bar
    B_{2}^{a_2}\big(\bar\xi_{\bar1}^- B_{\bar2}\big)^{\bar\jmath}
    \Big\rangle
    \\
    &\quad\, -\frac g 2\frac {\hat 1}{\hat 2(1\:3)}\Big\langle
    \xi_{1}^+{ }_i\:\big(\bar\xi_{\bar1}^- T^{a_2} \xi_{\bar3}^+
    \big)\bar \xi_{3}^{-\bar \jmath}\Big\rangle
  \end{split} \notag \\
  &= \frac{ig}2\frac{ i\sqrt2\: \hat 1}{(p_1^2-m^2)} \frac
  i{p_2^2}\frac{i\sqrt2\: \hat 3} {p_3^2-m^2}
  \left(2\times\frac{\{2\:3\}} {\hat2\hat
      3}+m^2\frac{\hat2}{(1\:2)\hat 3}\right)(T^{a_2})_i{}^{\bar
    \jmath},
  \label{eq:completeV-qpq-1}
  \\
  \begin{split} \Big\langle\alpha_{1}^-{ }_i\:\bar A^{a_2}_{2}\bar
    \alpha_{ 3}^{+\bar \jmath}\Big\rangle &= -\frac g{\sqrt 2}\frac
    {\hat 3}{(2\:3)}\Big\langle \big(B_{\bar2}\xi_{\bar3}^-\big)_i\bar
    B_{2}^{a_2}\bar \xi_{3}^{+\bar\jmath}\Big\rangle -\frac
    g{\sqrt2}\frac {\hat 1+\hat2}{(1\:2)}\Big\langle \xi_{1 }^-{ }_i\:
    \bar B_{2}^{a_2}\big(\bar\xi_{\bar1}^+ B_{\bar2}\big)^{\bar\jmath}
    \Big\rangle
    \\
    &\quad\, +\frac g 2\frac {\hat 3}{\hat 2(1\:3)}\Big\langle \xi_{1
    }^-{ }_i\:\big(\bar\xi_{\bar1}^+ T^{a_2} \xi_{\bar3}^-\big) \bar
    \xi_{3}^{+\bar\jmath}\Big\rangle
  \end{split} \notag
  \\
  &= -\frac{ig}2\frac{ i\sqrt2\: \hat 1}{(p_1^2-m^2)} \frac
  i{p_2^2}\frac{i\sqrt2\: \hat 3} {p_3^2-m^2}
  \left(2\times\frac{\{1\:2\}} {\hat1\hat
      2}+m^2\frac{\hat2}{(1\:2)\hat
      1}\right)(T^{a_2})_i{}^{\bar\jmath}\,.
  \label{eq:completeV-qpq-2}
\end{align}
By writing it this way, with the propagators made manifest, it is easy
to read off the amputated correlation function from these equations.
Notice that the second term in the big bracket of
(\ref{eq:completeV-qpq-1}) is just the same as the coefficient of the
first term in the parentheses of (\ref{eq:lcqcd-quark-m1-p}). As a
result, the first term in (\ref{eq:MHV-qpq-m}) is cancelled with the
contribution from second term in (\ref{eq:completeV-qpq-1}) after LSZ
procedure.  Similarly, the second term in (\ref{eq:MHV-qpq-m}) is
cancelled with the contribution from the second term in
(\ref{eq:completeV-qpq-2}). So the only terms left are the first terms
of (\ref{eq:completeV-qpq-1}) and (\ref{eq:completeV-qpq-2}). It is
worth noticing that this cancellation happens before we apply the LSZ
procedure. We conclude that this cancellation holds for all four
amplitudes $A(1_{\rm q}^{\pm}2^+3_{\bar {\rm q}}^{\pm})$.  Note that
it is impossible to send all three external particles on shell at the
same time in Minkowski signature, and these completion vertices
contribute to the off-shell amplitude.  If one goes to $(2,2)$
signature, one \emph{can} discuss the on-shell amplitude, and in this
case, one can easily verify that these completion vertices give zero
contribution using the on-shell identity
\begin{eqnarray}
  (1\:2)\{1\:2\}=(2\:3)\{2\:3\}=(1\:3)\{1\:3\}=-\frac {m^2}2\hat 2^2.
\end{eqnarray}
At tree level, the completion vertices make no contribution to the
higher point amplitudes. This is clear because these vertices can
survive LSZ reduction only by diverging, but by inspection they
generically diverge only for collinear momenta. Since in the higher
point vertices the kinematical constraints are insufficient to force
the on-shell momenta to be collinear, none of these terms survive the
on-shell limit.

Next, we turn to \eqref{eq:lcqcd-qbarpq} from the light-cone QCD
lagrangian, which in momentum space is
\begin{eqnarray}
  L^{\bar\psi+\psi}=g\left(\frac{\{1\:2\}}{\hat1\hat2} \bar
    \alpha_{\bar1}^+A_{\bar2}
    \alpha_{\bar3}^--\frac{\{2\:3\}}{\hat2\hat3}\bar\alpha_{\bar1}^-A_{\bar2}\alpha_{\bar3}^+\right).
\end{eqnarray}
It is clear that what this contributes to the amplitude is just what
is left in (\ref{eq:completeV-qpq-1}) and (\ref{eq:completeV-qpq-2})
(\ie\ their first terms), which we just calculated using the MHV
lagrangian. (The relative minus sign in (\ref{eq:completeV-qpq-1}) and
(\ref{eq:completeV-qpq-2}) comes from fermion statistics.)  From the
foregoing it is clear that this holds for all four amplitudes.

As a completion, we provide the four amplitudes $A(1_{\rm
  q}^{\pm}2^+3_{\bar {\rm q}}^{\pm})$:
\begin{eqnarray}
  &&\begin{aligned}
    A(1_{\rm q}^{+}2^+3_{\bar {\rm q}}^{-})=-ig\nu_1\nu_3
    \bigg[&2m^2\Big(\frac {\bar 1 \bar 2\tilde3}{\hat 2}-\frac{\bar1}{\hat 1}
    (3^--\hat 3)(\hat 1+3^{+})\Big)
    \\
    &-4\frac{1^{+}3^{+}\{1\:2\}}{\hat 1 \hat 2}
    (1^--\hat1)(3^--\hat3)\bigg]\,,
  \end{aligned}\label{eq:QCD-massive-ppm}
  \\
  &&\begin{aligned}
    A(1_{\rm q}^{-}2^+3_{\bar {\rm q}}^{+})=-ig\nu_1\nu_3
    \bigg[&2m^2\Big(\frac {\tilde 1 \bar 2\bar3}{\hat 2}-\frac{\bar3}{\hat 3}
    (1^--\hat 1)(\hat 3+1^{+})\Big)
    \\
    &+4 \frac{1^{+}3^{+}\{2\:3\}}{\hat 2 \hat 3}
    (1^--\hat1)(3^--\hat3)\bigg]\,,
  \end{aligned}\label{eq:QCD-massive-mpp}
  \\
  &&\begin{aligned}
    A(1_{\rm q}^{-}2^+3_{\bar {\rm q}}^{-})=-ig\nu_1\nu_3
    \bigg[&2\sqrt2m\Big(\frac {1^+ \bar 2\tilde3}{\hat 2}(1^--\hat
    1)+\frac{\tilde1\bar23^+}{\hat 2}(3^--\hat3)
    \\
    &-(1^--\hat 1)(3^--\hat3)(1^++3^{+})\Big)
    \bigg]\,,
  \end{aligned}\label{eq:QCD-massive-mpm}
  \\
  &&\begin{aligned}
    A(1_{\rm q}^{+}2^+3_{\bar {\rm q}}^{+})=-ig\nu_1\nu_3
    \bigg[&2\sqrt2m\Big(\frac {\bar1 3^+\{2\:3\}}{\hat 2\hat3}(3^--\hat
    3)-\frac{1^+\bar3\{1\:2\}}{\hat1\hat 2}(1^--\hat1)\Big)
    \\
    &+\sqrt2 m^3\frac{\bar1\hat 2\bar3}{\hat1\hat3}
    \bigg]\,.
  \end{aligned}\label{eq:QCD-massive-ppp}
\end{eqnarray}
Comparing with the results in the previous section one can also see
that flipping all external helicities corresponds to changing the
holomorphic variables to anti-holomorphic variables in the first two
amplitudes, together with an extra minus sign in the last two
amplitudes as we proved before.

\section{Conclusion}
\label{sec:conclusion}

In this paper, we extended the canonical MHV lagrangian formalism of
\cite{Mansfield:2005yd,Ettle:2006bw,Ettle:2007qc}, and thus proved
that we can obtain a canonical MHV lagrangian formalism for massless
QCD theory with ${\rm SU}(N_{\rm C})$ gauge symmetry.  We started with
massless QCD in the light-cone gauge with the non-dynamical field
components integrated out. By applying a canonical transformation to
the field variables, we obtain a lagrangian incorporating gluon-gluon
and quark-gluon interactions whose vertices must be proportional (up
to polarisation factors) to the MHV amplitudes in the literature.  For
completeness, this has been checked explicitly for amplitudes with two
quarks and two gluons, with four quarks, and with two quarks and three
gluons in the $(1_{\rm q}^+ 2^+ 3^- 4^+ 5_{\bar{\rm q}}^-)$
configuration.

The MHV QCD lagrangian we have found maintains `backward
compatibility' with the pure-gauge case found in
\cite{Mansfield:2005yd,Ettle:2006bw}. We note that the pure-gauge part
is preserved, as is the solution for $\gA$ in terms of $\gB$. In
contrast $\bar\gA$ acquires new terms in the new fermion fields
brought on by the requirement that the transformation is canonical. As
in the pure-gauge case, the explicit form of this transformation as a
series expansion has coefficients that have simple, holomorphic
expressions in the momenta. We summarise these results below:
\begin{align*}
  \gA_1 &= \sum_{n=2}^\infty \int_{2\cdots n} \Upsilon(1\cdots n)
  \gB_{\bar 2} \cdots \gB_{\bar n} \: (2\pi)^3 \delta^3({ \textstyle
    \sum_{i=1}^n \vec p_i
  })\\
  \begin{split}
    \bar\gA_1 &= \sum_{m=2}^\infty \sum_{s=2}^m \int_{2\cdots m}
    \frac{\hat s^2}{\hat1^2} \Upsilon(1 \cdots m) \gB_{\bar2} \cdots
    \bar\gB_{\bar s} \cdots \gB_{\bar m} \: (2\pi)^3
    \delta({\textstyle
      \sum_{i=1}^m \vec p_i}) \\
    &\quad + \frac{g^2}{\hat1^2 \sqrt 8} \sum_\pm \sum_{n=3}^\infty
    \sum_{j=1}^{n-2} \int_{2\cdots n} \left\{\begin{matrix}
        -\widehat{j+1} \\ \widehat{j+2}
      \end{matrix} \right\} \Upsilon(1 \cdots n) \\
    & \qquad \times \Biggl\{ \gB_{\bar 2} \cdots \gB_{\bar\jmath}
    \xi^\mp_{\overline{j+1}} \bar\xi^\pm_{\overline{j+2}}
    \gB_{\overline{j+3}} \cdots \gB_{\bar n} + \frac1{N_{\rm C}}
    \bar\xi^\pm_{\overline{j+2}} \gB_{\overline{j+3}} \cdots \gB_{\bar
      n} \gB_{\bar 2} \cdots \gB_{\bar\jmath} \xi^\mp_{\overline{j+1}}
    \Biggr\}
    \\
    &\qquad \times (2\pi)^3 \delta^3({\textstyle \sum_{i=1}^n \vec
      p_i}),
  \end{split} \\
  \bar\alpha^\pm_1 &=\bar\xi^\pm_1 + \sum_{n=3}^\infty \int_{2\cdots
    n} \left\{\begin{matrix} 1 \\ -\hat1/\hat2
    \end{matrix} \right\} \Upsilon(1\cdots n) \: \bar\xi^\pm_{\bar 2}
  \gB_{\bar 3} \cdots \gB_{\bar n}\: (2\pi)^3 \delta^3({ \textstyle
    \sum_{i=1}^n \vec p_i}), \\
  \alpha^\pm_1 &= \xi^\pm_1 + \sum_{n=3}^\infty \int_{2\cdots n}
  \left\{\begin{matrix} 1 \\ -\hat n/\hat1
    \end{matrix} \right\} \Upsilon(1\cdots n) \: \gB_{\bar 2} \cdots
  \gB_{\overline{n-1}} \xi^\pm_{\bar n}\: (2\pi)^3 \delta^3({
    \textstyle \sum_{i=1}^n \vec p_i}),
\end{align*}
where the stacked expressions in braces take their value in accordance
with the upper or lower choice of sign, and
\[
\Upsilon(1\cdots n) = (-i)^n \frac{\hat 1 \hat 3 \cdots \widehat{n-1}}
{(2\:3) \cdots (n-1,n)}.
\]

We have also begun the analysis of the MHV lagrangian for QCD with
massive quarks. We have observed that we can retain much of the
structure from the massless case, and although we defer a detailed
analysis of the amplitudes following from this lagrangian, we see that
the process results again in an infinite series of terms, but where
the chirality rather than the helicity content of each term is
constrained.  In the process, we derived explicitly the closed-form
expressions for the new vertices that arise in the massive MHV
lagrangian which are proportional to $m$ and $m^2$.  We cannot comment
here whether or not such an approach would yield a good computational
alternative to established techniques such as BCFW recursion relations
for massive particles
\cite{Badger:2005jv,Badger:2005zh,Schwinn:2006ca,Schwinn:2007ee,Ferrario:2006np,
  Ozeren:2006ft,Hall:2007mz}, but we remark that the important
question is whether it helps simplify the calculation of amplitudes
involving massive quarks at next-to-leading order and beyond.

As we found previously in \cite{Ettle:2007qc}, completion vertices,
arising from the terms in the transformation itself, are necessary to
recover (parts of) otherwise missing amplitudes.  We have verified
this in a simple case for massless QCD. They are of course necessary
for recovering the full off-shell theory \cite{Ettle:2007qc}, which we
showed in some example cases for massive QCD.  We expect these
completion vertices to be important at the loop level.

\section*{Acknowledgments}
The authors would like to thank Paul Mansfield for helpful discussions
at the start of this project. Tim, Zhiguang and James thank the STFC
for financial support.

\appendix

\section{Proofs of Field Transformation Coefficients}
\label{sec:seriesproofs}

In this appendix, we provide a proof for the expression
\eqref{eq:Sl-coeff} for $S^-$ and outline the proof of
\eqref{eq:K--coeff} for $K^-$, as given in section
\ref{ssec:series-solns}.

\subsection{$S^-$ coefficients}
\label{ssec:S-proof}
$S^-$ has coefficients defined by the recurrence relations
\eqref{eq:Srl-rr}.  Using \eqref{eq:Sl-coeff} to substitute for $S^-$,
the RHS of \eqref{eq:Srl-rr} becomes
\begin{equation}
  \label{eq:Sr-rr-RHS} 
  \begin{split} 
    & (-i)^n \frac{\hat 1 \hat 4 \cdots \widehat{n-1}}{(3\:4) \cdots
      (n\!-\!1,n)} \left\{ \frac{\hat 3}{(3,2\!+\!P_{3n})} +
      \sum_{j=3}^{n-1} \frac{(j,j\!+\!1)}{(j,2\!+\!P_{j,n})}
      \frac{\hat 2+\hat P_{j+1,n}}{(j\!+\!1,2\!+\!P_{j+1,n})} \right\}
    \\
    &= (-i)^n \frac{\hat1 \hat4 \cdots \widehat{n-1}}{(3\:4) \cdots
      (n\!-\!1,n)} \Biggl( x_3 + \sum_{j=3}^{n-1} y_j\Biggr)
  \end{split}
\end{equation}
where
\begin{equation}
  y_j = \frac{(j,j\!+\!1)}{(j,2\!+\!P_{jn})} \frac{\hat
    2 + \hat P_{j+1,n}}{(j\!+\!1,2\!+\!P_{j+1,n})}
  \quad\text{and}\quad
  x_j = \frac{\hat\jmath}{(j,2\!+\!P_{jn})}.
\end{equation}
Now notice that $x_j + y_j = x_{j+1}$, so the sum on the RHS of
\eqref{eq:Srl-rr} collapses to
\begin{equation}
  (-i)^n \frac{\hat1 \hat4 \cdots \widehat{n-1}}{(3\:4)
    \cdots (n\!-\!1, n)} x_n = S^-(12;3\cdots n),
\end{equation}
and the proof is complete.

\subsection{$K^-$ coefficients}
\label{ssec:K-proof}
Here we will outline the proof that expression \eqref{eq:K--coeff} for
$K^-$ satisfies the recurrence relation \eqref{eq:tildeKR}. As noted
in the main text, the proof is by induction on $n$. We treat the $n =
3$ case separately since it involves only the last term on the RHS of
\eqref{eq:tildeKR}.

For higher $n$, we substitute \eqref{eq:K--coeff} into the recurrence
relation. For each term on the RHS of \eqref{eq:tildeKR}, we can pull
out a factor of
\[
\frac{\hat 1 \cdots \hat n}{(1\:2)\cdots(n\:1)}
\]
leaving telescoping sums of the form
\begin{equation}
  \sum_{j=a}^b \hat P_{ij} \left(\frac{\widetilde{j+1}}{\widehat{j+1}}  - \frac{\tilde\jmath}{\hat\jmath}\right)
  = \hat P_{ib} \frac{\widetilde{b+1}}{\widehat{b+1}} - \hat P_{ia}
  \frac{\tilde a}{\hat a} - \tilde P_{a+1,b}.
  \label{eq:Delta-telescope-sum}
\end{equation}
(In the second term of \eqref{eq:tildeKR}, two such sums are nested.)
The cases noted below \eqref{eq:tildeKR}, where a sum is taken to
vanish if its upper limit is less than its lower limit, can be handled
consistently by defining the sum $P_{ij} = p_i + p_{i+1} + \cdots +
p_n + p_1 + \cdots + p_j$ when $j<i$. When this is so and we evaluate
the sums using \eqref{eq:Delta-telescope-sum}, we find that they
vanish in these particular conditions because of terms of the form
$P_{i,i-1}=0$.

To complete the proof, one simply evaluates the sums and does the
algebra while applying the conservation of momentum. This results in
an expression on the RHS of \eqref{eq:tildeKR} equal to the given for
$K^{-(j)}(1\cdots n)$ in \eqref{eq:K--coeff}.

\end{document}